\documentclass{article}

\usepackage{amsmath}
\usepackage{amsthm}
\usepackage{amsfonts}

\author{Thomas W. Cusick\thanks{Maxwell Bileschi and Daniel Padgett, undergraduate students supported by NSF CSUMS grant 0802994, contributed to this work.}\\
University at Buffalo, Department of Mathematics,\\
244 Mathematics Building, Buffalo, NY 14260\\
Email: cusick@buffalo.edu}
\title{Affine equivalence of cubic homogeneous rotation symmetric functions}
\date{}
\begin{document}

\newtheorem{theorem}{Theorem}[section]
\newtheorem{lemma}[theorem]{Lemma}
\newtheorem{proposition}[theorem]{Proposition}
\newtheorem{corollary}[theorem]{Corollary}

\newtheoremstyle{example}{\topsep}{\topsep}%
     {}
     {}
     {\bfseries}
     {.}
     {.5em}
     {\thmname{#1}\thmnumber{ #2}\thmnote{ #3}}

\theoremstyle{example}
\newtheorem{remark}[theorem]{Remark}
\newtheorem{example}[theorem]{Example}

\newcommand{\set}[1]{\left\{ #1\right\}}
\renewcommand{\vec}[1]{\textbf{#1}}
\newcommand{\integer}[1]{\left[#1\right]}
\newcommand{\Mod}[1]{\text{ Mod }#1}

\maketitle
NOTICE:  this is the author's version of a work that was accepted for publication in Information Sciences. Changes resulting
from the publishing process, such as peer review, editing, corrections, structural formatting, and other quality control mechanisms may not be reflected in this document.  Changes may have been made to this work since it was submitted for publication.  A
definitive version was subsequently published in Information Sciences 181 (2011), 5067-5083.  DOI 10.1016/j.ins.2011.07.002

\begin{abstract}
Homogeneous rotation symmetric Boolean functions have been extensively studied in recent years because of their applications in cryptography.  Little is known about the basic question of when two such functions are affine equivalent.  The simplest case of quadratic rotation symmetric functions which are generated by cyclic permutations of the variables in a single monomial was only settled in 2009.  This paper studies the much more complicated cubic case for such functions. A new concept of \emph{patterns} is introduced, by means of  which the structure of the smallest group $G_n$, whose action on the set of all such cubic functions in $n$ variables  gives the affine equivalence classes for these functions under permutation of the variables, is determined.  We conjecture that the equivalence classes are the same if all nonsingular affine transformations, not just permutations, are allowed.  Our method gives much more information about the equivalence classes; for example, in this paper we give a complete description of the equivalence classes when $n$ is a prime or a power of $3$. 
\end{abstract}

\section{Introduction}
Boolean functions have many applications in coding theory and
cryptography.  A detailed account of the latter applications can be found in
the book \cite{CBF}.  If we define $V_n$ to be the vector space of dimension $n$ over the
finite field  $GF(2) = \set{0, 1}$, then an $n$ variable Boolean function  $f(x_1, x_2, ..., x_n)
= f(\vec{x})$ is a map from $V_n$ to $GF(2)$.  Every Boolean function $f(\vec{x})$ has a unique
polynomial representation (usually called the \emph{algebraic normal form} \cite[p. 6]{CBF}),
and the degree of  $f$  is the degree of this polynomial.  A function of degree
$\leq 1$ is called affine, and if the constant term is 0 such a function is called linear. If every term in the algebraic normal form of $f$ has the same degree, then the function is \emph{homogeneous}.  All functions studied in this paper will be homogeneous.
We let  $B_n$ denote the set of all Boolean functions in $n$ variables, with addition
and multiplication done mod 2.

If we list the $2^n$ elements of $V_n$ as $v_0 = (0,\ldots,0), v_1 = (0,\ldots,0,1), \ldots$ in
lexicographic order, then the $2^n$-vector $(f(v_0), f(v_1),\ldots,f(v_{2^n - 1}))$ is
called the truth table of $f$.  The weight (also called Hamming weight)  $wt(f)$
of $f$ is defined to be the number of 1's in the truth table for $f$.  In many cryptographic uses of Boolean functions, it is important that the truth table of each function  $f$  has an equal number of 0's and 1's; in that case, we say that the function $f$  is balanced.

The distance $d(f, g)$ between two Boolean functions $f$ and $g$ is defined by

\begin{equation*}
d(f, g) = wt(f + g)
\end{equation*}

\noindent where the polynomial addition is done mod 2.  An important concept in
cryptography is the nonlinearity $N(f)$ defined by

\begin{equation*}
N(f) =  \min_{a\text{ affine}}  wt(f + a).
\end{equation*}

We say a Boolean function  $f(\vec{x})$ in $B_n$ 
is rotation symmetric if the algebraic normal form of the function is
unchanged by any cyclic permutation of the variables $x_1, x_2, \ldots, x_n$.  In recent years, rotation symmetric functions have proven to be very useful in several areas
of cryptography \cite[pp. 108 - 118]{CBF}.  This has led to many papers which study
different aspects of the theory of rotation symmetric functions (see the references in  \cite[pp. 108 - 118]{CBF}; some especially nice applications are in \cite{KMY}).

We say that two Boolean functions $f(\vec{x})$ and $g(\vec{x})$ in $B_n$ are \textit{affine equivalent}  if 
$g(\vec{x}) = f(A\vec{x} +  \vec{b})$, where $A$ is an $n$ by $n$ nonsingular matrix over the finite field $GF(2)$ and $b$ is an $n$-vector over $GF(2)$. We say $f(A\vec{x} +  \vec{b})$ is a \textit{nonsingular affine transformation} of $f(\vec{x})$.  It is easy to see that if $f$ and $g$ are affine equivalent, then $wt(f) = wt(g)$ and $N(f) = N(g)$.  We say that the weight and nonlinearity are \textit{affine invariants}.

One basic question is to decide when two Boolean functions $f(\vec{x})$ and $g(\vec{x})$ in
$B_n$ are affine equivalent. This question is nontrivial even for $n = 2$.  The next section is devoted to this quadratic case.

\section{Affine equivalence of quadratic rotation symmetric Boolean functions}
\label{sec2}

Before turning to the cubic functions, we look at what can be proved in the simpler quadratic case.  We shall consider only the 
simplest quadratic  functions $f$, namely those generated by cyclic permutations of the
variables in a single monomial. We shall call such functions \textit{monomial rotation symmetric functions}, or MRS functions, for brevity.  Thus any quadratic MRS function  
 $f(\vec{x})$ in $n$ variables can be written as

\begin{equation}
\label{MRSF}
f_{n,j}(\vec{x}) = x_1 x_j + x_2 x_{j+1} +...+ x_n x_{j-1}
\end{equation}

\noindent for some $j$ with  $2 \leq  j  \leq  \integer{\frac{n+1}{2}}$, or, in the special case when $n$ is even and $j = \frac{n}{2} + 1$, as

\begin{equation}
\label{SQF}
f_{n,\frac{n}{2} + 1}(\vec{x}) = x_1 x_j + x_2 x_{j+1} +...+x_{\frac{n}{2}} x_n.
\end{equation}

This latter function has only $\frac{n}{2}$ terms, whereas the functions in  \eqref{MRSF} have $n$ terms.  Because of this, we shall call the function   $f_{n,\frac{n}{2} + 1}(\vec{x})$  the \textit{short quadratic function} in $n$ variables.

Even in the quadratic case, it is necessary to consider restricted classes of functions, because the affine
equivalence problem for general functions is notoriously difficult.  Some work on restricted classes of cubic functions is in the papers \cite{CRB, Hou}. 

The basic theorem on affine equivalence of general quadratic Boolean functions
was proved by Dickson; his 1901 book on this and related topics has been reprinted
in \cite{Dick}.  A modern exposition of Dickson's work from a coding theory viewpoint is in \cite[pp. 438-442]{McWS}.

\begin{theorem}
\emph{(Dickson)}
\label{Dickson}
Suppose $f$ in $B_n$ has degree 2.  If  $f$  is balanced, then $f$ is affine equivalent to  $x_1x_2 + x_3x_4 + \ldots + x_{2k - 1}x_{2k} + x_{2k + 1}$ for some
$k \leq \frac{n - 1}{2}$.  If $f$ is not balanced, then $f$ is affine equivalent to 
$x_1x_2 + x_3x_4 + \ldots + x_{2k - 1}x_{2k} + b$ for some $k  \leq \frac{n}{2}$ and $b$ in $GF(2)$.
If $wt(f) < 2^{n - 1}$, then  $b = 0$.  If $wt(f) > 2^{n - 1}$, then  $b = 1$.
\end{theorem}

Given a function $f$ of degree 2, after we find the quadratic form in Theorem \ref{Dickson} which is equivalent to $f$ (unfortunately to do this is not trivial), it is easy to compute $wt(f)$ and $N(f)$.  The result is

\begin{lemma}
\label{lemone}
Suppose $g$ in $B_n$ has the form $\sum_{i=1}^k x_{2i - 1} x_{2i} + 
\sum_{i=2k + 1}^n a_i x_i$ with $k \leq \frac{n}{2}$.  Then $N(g) = 2^{n - 1} - 2^{n - k -1}$.  If all of the $a_i$ are 0, then $wt(g) = N(g)$; otherwise $wt(g) = 2^{n - 1}$, so $g$ is balanced.
\end{lemma}

\begin{proof}
Two different proofs appear in \cite[pp. 441-442]{McWS} and \cite[Lemma 5, p. 429]{Kim}.\renewcommand{\qedsymbol}{}%
\end{proof}

Our next lemma (well-known to experts in this area) follows from Theorem \ref{Dickson} and Lemma \ref{lemone}.

\begin{lemma}
\label{lem2}
Two quadratic functions $f$ and $g$ in $B_n$ are affine equivalent if and only if
$wt(f) = wt(g)$ and $N(f) = N(g)$.
\end{lemma}

\begin{remark}
For functions of degree $> 2$, it is not true that the affine invariants weight and nonlinearity suffice to determine the affine equivalence classes.  An example is 
$f_1(x) = x_1 x_4$ and $f_2(x) = x_1 x_2 x_3 + x_1 x_4$ in $B_4$.  These two functions both have weight and nonlinearity equal to 4, but they are not affine equivalent since they have different degrees.
\end{remark}

The weight and nonlinearity of the quadratic MRS functions $f_{n,2}$ were determined in \cite{PQ} and \cite[pp. 292-297]{CS02} (the latter paper supplied proofs for some cases not done in the former paper).  A much simpler proof of these results was given by Kim et al. in \cite[Lemma 7, p. 430]{Kim}.  Furthermore, in \cite[Theorem 8, p. 431]{Kim} the weight and nonlinearity of all of the MRS functions $f_{n,j}(\vec{x})$ was determined by using a new method.  Their work associates the permutation $\rho_{n,j}$
defined by

\begin{equation}
\label{eqn3}
\rho_{n,j}(i) \equiv  i + j - 1 \bmod{n}~~\text{for}~~j = 1, 2, \ldots, n
\end{equation}

\noindent with the function  $f_{n,j}(\vec{x})$  defined in \eqref{MRSF}. Note that this permutation is just a cyclic shift of the integers $1, 2, \ldots, n$. They prove the following theorem which determines the weight and nonlinearity of  $f_{n,j}$  \cite[Theorem 8 and Remark 10, p. 431]{Kim}.

\begin{theorem}
\emph{(Kim et al.)}
\label{thm2}
Suppose that the permutation $\rho_{n,j}$ associated with the function  $f_{n,j}$,  $2 \leq  j  \leq  \integer{\frac{n+1}{2}}$, has the disjoint cycle decomposition  $\tau_1 \tau_2 \ldots \tau_k$. Then the number of cycles is $k = \gcd(n, j - 1)$ and all the cycles have the same length $\frac{n}{k}$.  Also for  $2 \leq  j  \leq  \integer{\frac{n+1}{2}}$ we have

\begin{align*}
wt(f_{n,j}) &= N(f_{n,j}) =  2^{n - 1} - 2^{n/2 + k -1}                        &&\mbox{if }\frac{n}{k}\mbox{ is even}\\
wt(f_{n,j}) &= 2^{n - 1},  N(f_{n,j}) = 2^{n - 1} - 2^{(n + k)/2 -1}    &&\mbox{if }\frac{n}{k}\mbox{ is odd}
\end{align*}

For the short quadratic function,

\begin{equation*}
wt(f_{n,\frac{n}{2} + 1}) =  N(f _{n,\frac{n}{2} + 1}) = 2^{n - 1} - 2^{\frac{n}{2} - 1}
\end{equation*}
\end{theorem}

\begin{theorem}
\label{thm3}
The quadratic MRS functions  $f_{n,r}$  and  $f_{n,s}$  are affine equivalent if and only if 
$\gcd(n, r - 1) = \gcd(n, s - 1)$.
\end{theorem}
\begin{proof}
The "if" part follows from Lemma \ref{lem2} and Theorem \ref{thm2}.  The "only if" part follows since by Lemma \ref{lem2} the hypothesis of affine equivalence implies $wt(f_{n,r}) = wt(f_{n,s})$
and $N(f_{n,r}) = N(f_{n,s})$.  Then by Theorem \ref{thm2} $\gcd(n, r - 1) = \gcd(n, s - 1)$.%
\end{proof}

Theorem \ref{thm2} shows that it is easy to compute the weight and nonlinearity for any MRS quadratic function $f_{n,j}$.  We only need to find the integer $k = \gcd(n, j - 1)$.
This gives a quick way to find the equivalent form in Theorem \ref{Dickson}.

We now have enough to prove that in finding a nonsingular affine transformation
which maps one quadratic MRS function to another equivalent one, we need only look at permutations of variables, not arbitrary nonsingular affine transformations.

\begin{theorem}
\label{thm4}
If two quadratic MRS functions in $B_n$ are affine equivalent, then there is a permutation of the $n$ variables which gives the equivalence.
\end{theorem}
\begin{proof}
We need not consider the short function \eqref{SQF}, because it is easy to see that
the affine equivalence class for the short function has only one element.  Suppose
that the two functions  $f_{n,r}$  and  $f_{n,s}$  of form \eqref{MRSF} are affine equivalent.  It follows from Lemma \ref{lem2} that $wt(f_{n,r}) = wt(f_{n,s})$ and $N(f_{n,r}) = N(f_{n,s})$.  Hence Theorem \ref{thm2} implies that $\gcd(n, r - 1) = \gcd(n, s - 1)$;  we let $k$ denote this common value.
It follows from Theorem \ref{thm2} and the definition \eqref{eqn3} of the permutation $\rho_{n,j}$ that the
permutations $\rho_{n,r}$ and $\rho_{n,s}$ have cycle decompositions of form

\begin{equation*}
\prod_{i=1}^k (i, i + j - 1, i + 2(j - 1), \ldots, i + (\frac{n}{k} - 1)(j - 1)),
\end{equation*}

\noindent where j = r and s, respectively.  We use the notation

\begin{equation*}
C_{i,j} = (i, i + j - 1, i + 2(j - 1), ..., i + (\frac{n}{k} - 1)(j - 1)),  1 \leq i  \leq  k
\end{equation*}
for the k cycles in the product.

There are many ways to define a permutation $\xi$  such that  $\xi(f_{n,r}) = f_{n,s}$.
One natural way is to define  $\xi$  by  taking  $\xi(1)  =  1$  and  $\xi(r) = s$ (that is,
$\xi$ maps the leading term  $x_1 x_r$ of  $f_{n,r}$ to the leading term $x_1 x_s$ of  $f_{n,s}$).
Then we can extend  $\xi$  to every entry in the cycle $C_{1,r}$, using the rotation symmetry of the functions, to get

\begin{equation*}
\xi(1 + u(r - 1)) \equiv  1 + u(s - 1) \bmod{n},~~ 0  \leq  u  \leq  \frac{n}{k} - 1.
\end{equation*}
Extending this same pattern to the other cycles  $C_{i,r}$, the complete definition
of  $\xi$  is

\begin{equation}
\label{eqn4}
\xi(i + u(r - 1)) \equiv  i + u(s - 1) \bmod{n},~~ 0  \leq  u  \leq  \frac{n}{k} - 1,  1  \leq  i  \leq  k.
\end{equation}
Clearly  $\xi(C_{i,r})  =  C_{i,s}$  for  $1 \leq i  \leq  k$ and this proves the theorem.
\end{proof}
\begin{remark}
The proof of Theorem \ref{thm4} shows that if   $f_{n,r}$  and  $f_{n,s}$  of form \eqref{MRSF} are affine equivalent, then we can define a permutation $\xi$ which maps  $f_{n,r}$ to $f_{n,s}$
by choosing $\xi$ to map the pair $\set{1,r}$ to the pair $\set{a,b}$ in either order, where $x_a x_b$ is any one of the $n$ monomials in the representation \eqref{MRSF} of $f_{n,s}$. In this case, $\xi(C_{i,r})$ may map to a cycle whose entries are a permutation of the entries in $C_{j,s}$ for some $j  \neq  i$. In the proof of Theorem \ref{thm4}, the simplest choice $a = 1, b = s$ was made.
\end{remark}

\begin{example}
We take $n =  10$  and consider $f_{10,3}$  and  $f_{10,5}$  in $B_{10}$.  
These functions are affine equivalent by Theorem \ref{thm3}.  Following the proof of Theorem \ref{thm4}, we can define a natural permutation $\xi$ such that $\xi(f_{10,3}) = f_{10,5}$ by letting
$\xi(1) = 1, \xi(3) = 5, \xi(2) =  2, \xi(4) = 6$.  Completing the definition of $\xi$ using \eqref{eqn4} gives

\begin{equation*}
\xi((1,3,5,7,9)) = (1,5,9,3,7) ~\text{ and }~ \xi((2,4,6,8,10)) = (2,6,10,4,8).
\end{equation*}

\noindent Thus this map $\xi$ maps the two cycles of  $\rho_{10,3} = (1,3,5,7,9)(2,4,6,8,10)$ to the two cycles of $\rho_{10,5} = (1,5,9,3,7)(2,6,10,4,8)$.

We can define another permutation $\xi_1$ such that $\xi_1(f_{10,3}) = f_{10,5}$ by letting $\xi_1(1) = 6, \xi_1(3) = 2, \xi_1(2) =  1, \xi_1(4) = 5$.   Then the method in the proof of Theorem \ref{thm4} gives the full definition of $\xi_1$ as

\begin{equation*}
\xi_1((1,3,5,7,9)) = (6,2,8,4,10) ~\text{ and }~ \xi_1((2,4,6,8,10)) = (1,5,9,3,7).
\end{equation*}

\noindent In this case $\xi_1$ maps the cycle $(2,4,6,8,10)$ in $\rho_{10,3}$ to the cycle $(1,5,9,3,7)$ in
 $\rho_{10,5}$, but $\xi_1$ maps the cycle $(1,3,5,7,9)$ in $\rho_{10,3}$ to a cycle $(6,2,8,4,10)$
in which the order of the integers in the corresponding cycle $(2,6,10,4,8)$ in  $\rho_{10,5}$ is permuted.
\end{example}

\begin{remark}
\label{rem3}
It is easy to see that we cannot extend Theorem \ref{thm4} to assert that if two quadratic MRS functions in $B_n$ are affine equivalent, then only permutations will give the equivalence.  For example, the function $f_{4,2}(\vec{x})$ (using the notation \eqref{MRSF}) in $B_4$ is affine equivalent to itself by the nonsingular nonpermutation map

\begin{align*}
y_1 &= x_1+x_2+x_3, &y_2 &= x_2+x_3+x_4,\\
y_3 &= x_1+x_3+x_4, &y_4 &= x_1+x_2+x_4,
\end{align*}

\noindent under which $f_{4,2}(\vec{x}) = f_{4,2}(\vec{y})$.  If we go up to 8 variables, then we can find an example of a quadratic MRS function which is affine equivalent to a different quadratic MRS function by a nonpermutation map.  We can take $f_{8,2}(\vec{x})$ and define the nonpermutation map by

\begin{align*}
x_1 &= w_2+w_4+w_7, &x_2 &= w_5+w_7+w_8, &x_3 &= w_4+w_7+w_8,\\ 
x_4 &= w_3+w_7+w_8, &x_5 &= w_4+w_6+w_7, &x_6 &= w_1+w_7+w_8,\\
x_7 &= w_7, &x_8 &= w_8.&&
\end{align*}

\noindent Now computation gives   $f_{8,2}(\vec{x}) =  f_{8,4}(\vec{w})$.
\end{remark}

\begin{remark}
\label{rem4}
It is also easy to see that there exist affine equivalent quadratic homogeneous functions which cannot be shown to be equivalent by any permutation of variables. We simply drop the hypothesis in Theorem \ref{thm4} that the two functions are rotation symmetric.  An example is $f_{4,2}(\vec{x})$ in $B_4$ and  $g(\vec{x}) = x_1x_2$ in $B_4$. These functions are easily seen to be affine equivalent by Theorem \ref{Dickson} or Lemma \ref{lem2}, but no permutation of variables can give this equivalence, since any permutation applied to a function preserves the number of variables which actually appear in that function.
\end{remark}

\section{Affine equivalence for cubic rotation symmetric Boolean functions}

Almost nothing is in the literature concerning affine equivalence for cubic
rotation symmetric Boolean functions.  We shall consider the simplest of
such functions $f$, namely those generated by cyclic permutations of the
variables in a single monomial.  These are the cubic monomial rotation symmetric (MRS) functions, in the terminology of Section \ref{sec2}.  Thus for some  $j$ and $k$, $1<j<k$, we have

\begin{equation}
\label{eqn5}
f(\vec{x}) = x_1 x_j x_k + x_2 x_{j+1} x_{k+1} +...+ x_n x_{j-1}x_{k-1}.
\end{equation}

We shall use the notation $(1,j,k)$  for the function $f(\vec{x})$ in \eqref{eqn5}, no matter how
the terms on the right-hand side are written (so the order of the terms,
and of the 3 variables in each term, does not matter).  If $(1,j,k)$ is 
written in the form \eqref{eqn5} (so the first subscripts in the $n$ terms are $1, 2, \ldots, n$
in order,  and the other two subscripts in order each give cyclic permutations
of $1, 2, \ldots, n$, as shown), we say $f$ is written in standard form.  Note we do not
require $j < k$, so there are two ways to write $f(\vec{x})$ in standard form.  If we specify the representation of $f(\vec{x})$ ( $(1,j,k)$ or $(1,k,j)$ ), then the standard form is unique.
Clearly each subscript $j$, $1 \leq j \leq n$, appears in exactly 3 of the terms
in any representation of $f(\vec{x})$; we shall call these three terms the $j$-terms
of $f$.  We shall use the notation

\begin{equation}
\label{eqn6}
[i, j, k]  =  x_i x_j x_k
\end{equation}

\noindent as shorthand for the monomial on the right-hand side;  note that the order of
the variables matters,  so, for example, the 6 permutations of $i, j, k$ give
6 different (but equal) representations of form \eqref{eqn6} for the same monomial   $x_i x_j x_k$.

If $n$ is divisible by 3, then the function $(1, \frac{n}{3} + 1, \frac{2n}{3} + 1)$ is exceptional because then the representation \eqref{eqn5} has only $\frac{n}{3}$ distinct terms, because the three $j$-terms
for any $j$ are all the same, apart from the order of their factors.  Thus for $n \equiv 0 \bmod{3}$
the representation \eqref{eqn5} reduces to a sum of only $\frac{n}{3}$  terms.  Because of this, we
shall call  $(1, \frac{n}{3} + 1, \frac{2n}{3} + 1)$ the \textit{short cubic function} in $n$ variables.

In order to study the affine equivalence classes for the functions $(1,j,k)$, we need to be able to identify all of the distinct functions  $(1,j,k)$.  We define

\begin{align*}
D_n = \{(1,j,k):~ &j < k \leq n, \text{ and every function (1,j,k) is represented by}\\
                 &\text{the triple 1,j,k  with least j, and given that, with least k}\}.
\end{align*}

\noindent Every cubic monomial rotation symmetric function $f$ is equal to exactly one
function $(1,j,k)$ in $D_n$, but of course $f$ is also equal to $(1,p,q)$, where $[1,p,q]$ is either of  the other two 1-terms in $(1,j,k)$.

Clearly we can determine $D_n$ by making a list of all of the functions
$(1,j,k)$ with $1< j < k \leq n$ in lexicographic order and standard form,
and then crossing out any function in the list which has a 1-term
appearing in any earlier function in the list.  The number of distinct
functions which remain after this is given in the following lemma (as usual, $|S|$ denotes the number of elements in the set $S$).

\begin{lemma}
\label{lem3}
If $n \equiv 0 \bmod{3}$, then $|D_n| = (n^2 - 3n + 6)/6$. Otherwise, $|D_n| = (n^2 - 3n + 2)/6$.
\end{lemma}
\begin{proof}
An equivalent formula was first computed by St\u{a}nic\u{a} and Maitra \cite[p. 302]{SM}. 
A direct counting proof is also possible. The "extra" function when $n \equiv 0 \bmod{3}$ is the short function $(1, \frac{n}{3} + 1, \frac{2n}{3} + 1)$,
which is the last function produced when $D_n$ is determined by the method above.
\end{proof}

We define the notion of \textit{pattern} for any term $[i,j,k]$.  The pattern of
$[i,j,k]$ is the integer vector 

\begin{equation}
\label{eqn7}
( j - i \Mod{n};  k - i \Mod{n};  k - j \Mod{n}).
\end{equation}
The semicolons in \eqref{eqn7} distinguish a pattern from a function $(i,j,k)$.  
Throughout the paper the "capital mod" notation  $a \Mod{n}$ means the unique integer $b$ in 
$\set{1,2,\ldots,n}$  such that $b \equiv a \bmod{n}$.  When the modulus $n$ is clear, we shall
omit the $\Mod{n}$ in the notation \eqref{eqn7}.  Every term $[i,j,k]$ has 6 patterns $(a; b; c)$,
one for each of the orderings of the triple $i, j, k$.

\begin{lemma}
\label{lem4}
Each function $(1,j,k)$ in standard form has a unique pattern
$(j - 1 \Mod{n};  k - 1 \Mod{n};  k - j \Mod{n})$, which is the same for all of the $n$
terms $[u,v,w]$ in the standard form of the function.
\end{lemma}
\begin{proof}
This is obvious since in the standard form \eqref{eqn5} the subscripts in each term
are obtained by adding 1 to each of the corresponding subscripts in the preceding term.%
\end{proof}

\begin{lemma}
\label{lem5}
Suppose $(1,j,k)$ in standard form and $(1,p,q)$ are cubic monomial rotation
symmetric functions in $n$ variables.  If  $\mu( (1,j,k) ) = (1,p,q)$ for some permutation $\mu$ of the $n$ variables, then all of the terms 

\begin{equation}
\label{eqn8}
[\mu(i), \mu(i+j-1), \mu(i+k-1)],~ 1  \leq  i  \leq  n
\end{equation}

\noindent can be rearranged to give a
standard form of the function $(1, p, q)$. All of these rearranged terms will have the same pattern.
\end{lemma}
\begin{proof}
We can order the terms in \eqref{eqn8}, permuting
their entries as necessary, to get the function $(1, p, q)$ in standard form.
Then Lemma \ref{lem5} follows from Lemma \ref{lem4}.
\end{proof}

We say a permutation $\sigma$ of the $n$ variables in a cubic function preserves rotation
symmetry if, given any cubic MRS function $f$ in $B_n$, $\sigma(f)$ is also rotation
symmetric. Our next theorem shows that if two cubic MRS functions in $B_n$ are affine
equivalent via a permutation of variables which preserves rotation symmetry, then there is a
computationally efficient method to find such a permutation, even one with the extra property that
the permutation fixes 1.  We conjecture that there is no loss of generality in considering only affine maps which are permutations (see Remark \ref{rem5} below).  Furthermore, in applications using rotation symmetric functions, functions which do not have rotation symmetry are usually of no interest, so the permutations which preserve rotation symmetry are the only important ones.

Before stating the theorem, it is useful to have a characterization of the permutations which preserve rotation symmetry. The next lemma gives this; note that the characterization is equivalent to \eqref{eqn9A} in the theorem below. There is no loss of generality in taking $n > 4$ in the next lemma and theorem, since the cases for smaller $n$ are trivial.

\begin{lemma}
\label{lem6}
A permutation $\mu$ preserves rotation symmetry for cubic MRS functions in $n > 4$ variables if and only if 

\begin{equation}
\label{eqn8A}
\mu(i) = (i - 1)(\mu(2) - 1) +1~ \rm{Mod}~n, ~1 \leq i \leq n.
\end{equation}

\end{lemma}
\begin{proof}
We note there is no loss of generality in assuming $\mu(1) = 1$. It is trivial that \eqref{eqn8A}
implies that $\mu$ preserves rotation symmetry; so we assume that $\mu$ preserves rotation symmetry and we shall prove \eqref{eqn8A}.  Throughout the proof, $\equiv$ will always mean congruence mod $n$.

\indent Suppose $\mu((1,2,3)) = (1,p,q)$.  We will write $(1,p,q)$ in the standard form which contains the term $\mu([2,3,4]) = [p,q,x]$.
Then, by Lemma \ref{lem5}, $[p,q,x]$ has the same pattern as some rearrangement of  $[1,p,q]$ and we want to determine the value of x.\\
\indent We know that there are six possible patterns for the monomial $[1,p,q]$, and these patterns are the six
\textbf{ordered} triples in the following list:
\begin{itemize}
\item[1] $(p-1; q-1; q-p)$ from $[1,p,q]$
\item[2] $(q-1; p-1; p-q)$ from $[1,q,p]$
\item[3] $(1-p; q-p; q-1)$ from $[p,1,q]$
\item[4] $(q-p; 1-p; 1-q)$ from $[p,q,1]$
\item[5] $(p-q; 1-q; 1-p)$ from $[q,p,1]$
\item[6] $(1-q; p-q; p-1)$ from $[q,1,p]$
\end{itemize}

The pattern of $[p,q,x]$ must be one of these six patterns, and in order to determine $x$ we test the six cases in sequence. We have\\
\\
Pattern of $[1,p,q] = (p-1;q-1;q-p)$ and\\
\\
Pattern of $[p,q,x] = (q-p;x-p;x-q)$, where $x = \mu(4)$.\\
\\
For the first case, assume $q-p \equiv p-1, \implies q\equiv2p-1$.\\
Then $x-p \equiv q-1, \implies x-p \equiv 2p-2 \implies x\equiv3p-2$.\\
Then $q-p \equiv 2p-1-p \equiv p-1$ should also be $\equiv x-q \equiv 3p-2-2p+1 \equiv p-1.$ This is true,
so $x\equiv3p-2$ or $x \equiv 3(p-1)+1$ is a possibility.\\
Next assume $q-p\equiv q-1$.  Then $p=1$, which is false.\\
Next assume $q-p\equiv 1-p$.  Then $q=1$, which is false.\\
Next assume $(q-p; x-p; x-q)$ = $(q-p; 1-p; 1-q)$. Then $1-p\equiv x-p$ and $1-q\equiv x-q$ so $x=1$.  This can only happen when $n=3$.\\
Next assume $q-p \equiv p-q$.  Then $2(q-p)\equiv 0$, so $2p \equiv 2q \pmod{n}$.\\
\hspace*{1em} Then, $x-p \equiv 1-q \implies x \equiv 1-q+p$, and $2x \equiv 2-2q+2p \equiv 2$.\\
\hspace*{1em} Also, $x-q \equiv 1-p \implies x \equiv 1-p+q \equiv 1-q+p$, as above.  So\\  
\hspace*{1em} $x \equiv 1-p+q$ is possible.\\
Lastly, assume $q-p\equiv 1-q$.  Then $2q \equiv 1+p, \implies p \equiv 2q-1$.\\
\hspace*{1em} Then, $x-p \equiv p-q \equiv 2q-1-q \equiv q-1$\\
\hspace*{2em} and $x - p = x-(2q-1) \implies x = (q-1)+(2q-1).$\\
\hspace*{2em} Thus $x \equiv 3q-2.$\\
\hspace*{1em} Then, $x-q \equiv p-1 \equiv 2q-2$ and\\
\hspace*{1em} $x - q \equiv (3q-2)-(q) \equiv 2q-2$ are consistent,   so $x = 3q-2$ is a possibility.\\ 
\\
Summarizing, we have  $x \in \{1-p+q, 3(q-1)+1, 3(p-1)+1\}$. We shall prove only the third choice for $x$ is valid.\\
\\
Now we consider these three possible choices for x in the next term $ [q,x,y]$ in $\mu((1,2,3))$.\\
\\
First suppose $x \equiv 1-p+q$, so we also have $2p \equiv 2q$ from the work above.  Then:\\
Pattern of $[q,x,y] = (1-p; y-q; y-1+p-q)$, where $y = \mu(5)$.\\
\\
First assume $1-p \equiv p-1$.  Then $2p \equiv 2 \pmod{n}$.\\
\hspace*{1em} Then $y-q\equiv q-1 \implies y \equiv 2q-1 \equiv 2-1 \equiv 1$.\\
\hspace*{1em} Then $\mu(1) = \mu(5) \implies 1 \equiv 5  \implies 0 \equiv 4  \implies n=4$.\\
Next assume $1-p=q-1$.  Then $q+p\equiv 2 \pmod{n}$.\\
\hspace*{1em} Then $y-q=p-1 \implies y=p+q-1 = 2-1 = 1$.\\
\hspace*{1em} Then $\mu(1) = \mu(5) \implies 1 \equiv 5  \implies n=4$.\\
Next assume $1-p \equiv 1-p$.\\
\hspace*{1em} Then $y-q \equiv q-p \implies y\equiv 2q-p \equiv 2p-p = p$.\\
\hspace*{1em} Then $\mu(2) = \mu(5) \implies n=3$.\\
Next assume $1-p \equiv q-p$.  Then $q=1$, impossible.\\
Next assume $1-p \equiv p-q$.  Then $1\equiv 2p-q \equiv 2q-q \equiv q$, so $q=1$, impossible.\\
Next assume $1-p \equiv 1-q$.  Then $p=q$, impossible.\\
\\
So we cannot get the next term and thus $x \equiv 1-p+q$ is not a valid choice.\\
\\
Next suppose $x \equiv 3(q-1)+1$, so also $p \equiv 2q - 1$ from the work above.  Then:\\
Pattern of $[q,x,y] = (2(q-1); y-q; y-3q+2)$.\\
\\
First assume $2(q-1) \equiv p-1$.\\
Then $p\equiv 2(q-1)+1 \equiv 2q-1$.\\
Then, $y-q \equiv q-1 \implies y\equiv 2q-1$.\\
Also, $q-p \equiv y-3q+2$\\
$\Leftrightarrow q-2(q-1)-1 \equiv 2q-1-3q+2$.\\
$\Leftrightarrow -q+1 \equiv -q+1$, which is true.
\\
So $y\equiv 2q-1\equiv p$ works numerically.  However, then $\mu(2) = p = y =\mu(5) \implies n=3$.\\
Next assume $2(q-1) \equiv q-1$.  Then $q=1$, impossible.\\
Next assume $2(q-1) \equiv 1-p \equiv 1-(2q-1) \equiv -(2q-2) \equiv -2(q-1)$.  Then $4(q-1)\equiv 0 \implies 4q\equiv 4 $.\\
\hspace*{1em} Then $p\equiv 2q-1 \equiv 4q-1-2q \equiv 3-2q$.\\
\hspace*{1em} Then $y-q \equiv q-p \equiv q-3+2q \equiv 3q-3$.\\
\hspace*{1em} So $y= 4q-3 \equiv 4-3 \equiv 1$.\\
\hspace*{1em} Then $\mu(1) = \mu(5) \implies n=4$.\\
Next assume $2(q-1) \equiv q-p \equiv q-(2q-1)$.  Then $2(q-1) \equiv -q+1,\implies 3(q-1)\equiv 0 \implies 3q\equiv 3$.\\
\hspace*{1em} Then $p \equiv 2q-1 \implies p \equiv 3q-1-q \equiv 3-1-q \equiv 2-q$.\\
\hspace*{1em} Then $y-q \equiv 1-p \implies y-q \equiv 1-(2-q) \equiv q-1$.\\
\hspace*{1em} So $y= 2q-1\equiv p$, so $\mu(2) = \mu(5) \implies n=3$.\\
Next assume $2(q-1) \equiv p-q \equiv (2q-1)-q \equiv q-1$.  Then $q-1\equiv0 \implies q=1$, impossible.\\
Next assume $2(q-1) \equiv 1-q$.  Then $3(q-1)\equiv 0 \implies 3q\equiv 3$.\\
\hspace*{1em} Then $y-q= p-q \equiv 2q-1-q \equiv q-1$, so $y\equiv 2q-1\equiv p$.\\
\hspace*{1em} Then $\mu(2) = \mu(5) \implies n=3$.\\
\\
So we cannot get the next term and thus $x = 3(q-1)+1$ is not a valid choice.\\
\\
Thus the only valid choice is $x\equiv 3(p-1)+1$, and so we also have $q\equiv 2p-1$ from the work above. Hence for $n > 4,~ \mu((1, 2, 3)) = (1, p, q) \implies \mu(2) = p,~ \mu(3) =
 q \equiv 2p - $1 and $\mu(4) \equiv 3p - 2.$  
\\
\indent We now wish to show by induction that 
$$ \mu(i) \equiv (i-1)(p-1)+1 \equiv (i-1)(\mu(2)-1)+1, $$
which will give \eqref{eqn8A}.
We already proved this for $i \in \{1,2,3,4\}$ as a base case.\\
Assume true for some $i \geq 4$, and for $i-1, i-2, i-3$.\\
Then $\mu([i-1, i, i+1]) = [(i-2)(p-1)+1, (i-1)(p-1)+1, x]$ and we need to determine the value of $x$.\\
\\
The pattern for this term is $(p-1; x - (i-2)p + (i-3); x - (i-1)p + (i-2) )$.\\
\\
First assume the pattern is $(p-1; q-1; q-p) = (p-1; 2p-2; p-1)$\\
$\Leftrightarrow x - (i-2)p + (i-3) \equiv 2p-2 \implies x= (i-2+2)p - (2+i-3) \equiv ip - (i-1).$
So $x = ip - (i-1) = (i+1-1)(p-1) + 1$ works.\\
Next assume $p-1 \equiv q-1$, then $q=p$, impossible.\\
Next assume $p-1 \equiv 1-p$.  Then $2p \equiv 2$.\\
\hspace*{1em} Then $x - (i-2)p + (i-3) \equiv q-p \equiv 2p-1-p \equiv p-1 \implies x \equiv (i-2+1)p - (i-3+1) \equiv (i-1)p - (i-2)$.\\
\hspace*{1em} Then $\mu(i) = \mu(i+1) \implies n=1$, impossible.\\
Next assume $p-1 \equiv q-p \equiv 2p-1-p \equiv p-1$.  Clearly true.  Then: \\
\hspace*{1em} $x - (i-2)p + (i-3) \equiv 1-p \implies x \equiv (i-2-1)p - (i-3-1) \equiv (i-3)p - (i-4).$
\hspace*{1em} Then $\mu(i+1) = \mu(i-2) \implies i+1 \equiv i-2  \implies 3 \equiv 0  \implies n=3$.\\
Next assume $p-1 \equiv p-q$.  Then $q=1$, impossible.\\
Next assume $p-1 \equiv 1-q \equiv 1-(2p-1) \equiv 2-2p$.  Then $3p-3\equiv0 \implies 3p 
\equiv 3$.\\
\hspace*{1em} Then $x - (i-2)p + (i-3) \equiv p-q \equiv p-(2p-1) \equiv 1-p$\\
\hspace*{1em} $\implies x \equiv (i-2-1)p - (i-3-1) \equiv (i-3)p - (i-4)\\ 
\hspace*{1em} \implies \mu(i+1) = \mu(i-2) \implies n=3$.\\
\\
Thus only the first case gives the value of $x$ and the proof by induction of \eqref{eqn8A} is complete.
\end{proof}

\begin{theorem}
\label{thm5}
Suppose $(1,j,k)$ in standard form and $(1,p,q)$ are cubic monomial rotation
symmetric functions in $n > 4$ variables.  If  $\mu( (1,j,k) ) = (1,p,q)$ for some
permutation $\mu$ of the $n$ variables which preserves rotation symmetry, then there exists a 
permutation $\sigma$
such that $\sigma( (1,j,k) ) = (1,p,q)$, $\sigma( [1,j,k] ) = [1,p_i,q_i]$ and $\sigma(1)=1$,
where  $[1,p_i,q_i] ~ (1 \leq i \leq 3)$ is one of the three 1-terms in $(1,p,q)$.  The
pattern of the term $[1,\sigma(j),\sigma(k)]$  in $\sigma( (1,j,k) )$ is

\begin{equation}
\label{eqn9}
(\sigma(2) - 1)( j - 1;  k - 1;  k - j),
\end{equation}

\noindent where $\gcd(\sigma(2) - 1, n) = 1$. Furthermore, $\sigma$ satisfies

\begin{equation}
\label{eqn9A}
\sigma(i) = (i-1)(\sigma(2)-1)+1~  \rm{Mod}~{n}, ~1 \leq i \leq n.
\end{equation}

\end{theorem}
\begin{proof}
We may assume without loss of generality that $(1,j,k)$ and $(1,p,q)$ are in $D_n$. Suppose  $\mu(v) = 1, 1  \leq v  \leq  n$.  Define the permutation $\delta$ by

\begin{equation*}
\delta(w) =  v + w - 1 \Mod{n}.
\end{equation*}
Since $\delta$ is a cyclic shift of  $1, 2, \ldots, n$, we have $\delta( (1,j,k) ) = (1,j,k)$.
Obviously  $\mu(\delta(1)) = \mu(v) = 1$, so we can take $\sigma = \mu \delta$.  Since
$\sigma( (1,j,k) ) = (1,p,q)$ and $\sigma(1) = 1$, we must have $\sigma([1,j,k]) = [1,p_i,q_i]$
where $[1,p_i,q_i]$ ($i = 1, 2$ or $3$) is one of the three 1-terms in $(1,p,q)$.

Now \eqref{eqn9A} follows from Lemma \ref{lem6}.  Next consider the pattern $(\sigma(j)-1; \sigma(k)-1; \sigma(k)-\sigma(j))$ of the term   

\begin{equation}
\label{eqn10}
\sigma( [1,j,k] ) = [1, \sigma(j), \sigma(k)].
\end{equation}
By Lemmas \ref{lem5} and \ref{lem6}, the term $[\sigma(2), \sigma(j + 1), \sigma(k + 1)]$  in  $\sigma( (1,j,k) ) = (1,p,q)$ must have the same pattern as the term in \eqref{eqn10}, so we have

\begin{equation*}
[\sigma(2), \sigma(j + 1), \sigma(k + 1)] = 
                          [1 + \sigma(2)-1, \sigma(j) + \sigma(2)-1, \sigma(k) + \sigma(2)-1].
\end{equation*}
Similarly, all of the terms $T_i = [\sigma(i), \sigma(j + i - 1), \sigma(k + i - 1)]$  in
$(1,p,q)$  for  $i = 1, 2, \ldots, n$ must satisfy

\begin{align}
\label{eqn11}
&[\sigma(i), \sigma(j + i - 1), \sigma(k + i - 1) ]  = \\
\notag&\hspace{2em} [1 +(i-1)( \sigma(2)-1), \sigma(j) +(i-1)( \sigma(2)-1), \sigma(k) +(i-1)( \sigma(2)-1)].
\end{align}
Thus $T_i$ is obtained from $T_{i-1}$ by adding   $\sigma(2)-1$ to each entry in $T_{i-1}$.
This is equivalent to \eqref{eqn9A}.  Also, this shows (take $i = j$ or $k$, respectively, in \eqref{eqn11})

\begin{equation}
\label{eqn12}
(j-1) ( \sigma(2)-1)  =   \sigma(j)-1  \Mod{n}
\end{equation}

\noindent and

\begin{equation}
\label{eqn13}
(k-1) ( \sigma(2)-1)  =   \sigma(k)-1  \Mod{n}.
\end{equation}
Subtracting \eqref{eqn12} from \eqref{eqn13} gives

\begin{equation}
\label{eqn14}
(k-j) ( \sigma(2)-1)  =   \sigma(k)-\sigma(j)  \Mod{n}.
\end{equation}
Together, \eqref{eqn12}, \eqref{eqn13}  and \eqref{eqn14} show

\begin{equation*}
(\sigma(j)-1; \sigma(k)-1; \sigma(k)-\sigma(j)) = ( \sigma(2)-1)(j-1; k-1; k-j),
\end{equation*}

\noindent that is, the pattern of $[1,\sigma(j),\sigma(k)]$  is \eqref{eqn9}, as stated in the Theorem.

From \eqref{eqn11} and the fact that $\sigma(i)$ must take on all values $1, 2, \ldots, n$ for
$1  \leq  i  \leq  n$, we see that

\begin{equation*}
\set{1 + (i-1)(\sigma(2) - 1):  1  \leq  i  \leq  n}  =  \set{1, 2, \ldots, n}.
\end{equation*}
Therefore  $\gcd(\sigma(2) - 1, n) = 1$ and the proof is complete.
\end{proof}

\begin{example}
\label{ex2}
We take $n = 8$, so that

\begin{equation*}
D_8 = \set{(1,2,3), (1,2,4), (1,2,5), (1,2,6), (1,2,7), (1,3,5), (1,3,6)}.
\end{equation*}

If  $\mu = (1,4,3,6,5,8,7,2)$ (we represent permutations as products of disjoint cycles), then 

\begin{align*}
\mu( (1,3,6) ) = &\mu([1,3,6] + [2,4,7] + \ldots + [7,1,4] + [8,2,5]) = \\
                 &([4,6,5] + [1,3,2] + \ldots + [2,4,3] + [7,1,8]) = (1, 3, 2).
\end{align*}
This gives $v = 2$ in the proof of Theorem \ref{thm5}, so $\delta(w) = w + 1 \bmod{8}$ and
$\sigma = \mu \delta = (1)(3)(5)(7)(2,6)(4,8)$.  Thus $\sigma( [1,3,6] ) = [1,3,2]$ and

\begin{align*}
&\sigma([1,3,6] + [2,4,7]+ \ldots + [7,1,4] + [8,2,5]) = \\
&         \hspace{1in}[1,3,2] + [6,8,7] + \ldots + [7,1,8] + [4,6,5] = (1,3,2).
\end{align*}
The pattern of $[1,3,6]$ is $(2; 5; 3)$,  $\sigma(2) - 1 = 5$ and the pattern of $[1,3,2]$
 is $(2; 1; 7) = 5(2; 5; 3)$, in accordance with Theorem \ref{thm5}. Notice we need the standard form generated by the term $[1,3,2]$, even though of course the functions $(1,2,3)$ and $(1,3,2)$ are the same. 
\end{example}

Let $\sigma_{\tau,n} = \sigma_\tau$ denote the permutation defined by

\begin{equation*}
\sigma_\tau(i) = (i-1)\tau+1\Mod{n}~~\text{for}~~i=1,2,\ldots, n, 
\end{equation*}
where we assume 
\begin{equation*}
\gcd(\tau,n)=\gcd(\sigma_\tau(2)-1,n)=1.
\end{equation*}
Then we have
\begin{equation*}
 \gcd(\sigma_\tau(j)-1,n)=\gcd((j-1)\tau, n)=1 \mbox{ if and only if } \gcd(j-1,n)=1.
\end{equation*}
Since $\sigma_{\tau}\sigma_{\delta}=\sigma_{\delta}\sigma_{\tau}=\sigma_{\tau \delta}$ for any $\delta$ with $\gcd(\delta,n)=1$, we see that 
$G_{n}$ defined by
\begin{equation*}
 G_{n}=\{\sigma_{\tau,n}:\gcd(\tau,n)=1\}
\end{equation*}
is a group with the group operation of permutation composition.

\begin{theorem}
\label{thm6}
 The group $G_{n}$ is isomorphic to the group $U_n$ of units of $\mathbb{Z}_{n}^*$ given by

\begin{equation*}
 U_n=\{k:\gcd(k,n)=1\}
\end{equation*}
with group operation multiplication mod n.

\end{theorem}

\begin{proof}
 The bijection $\sigma_\tau \leftrightarrow \tau$ is a group isomorphism.
\end{proof}

\begin{theorem}
\label{thm7}

 The group $G_n$ acts on the set
\begin{equation*}
 C_n=\{cubic\;MRS\;functions\;f(\vec{x})\;in\;n\;variables\}
\end{equation*}
by the definition
\begin{equation}
\label{eqn15}
 \sigma_{\tau,n}(f(\vec{x})) = \sigma_{\tau,n}((1,j,k))
\end{equation}
where $f(\vec{x})$ has the unique standard form $(1,j,k)$ in $D_n$. The orbits for this group action are exactly the affine equivalence classes for $C_n$ under permutations which preserve rotation symmetry. 
\end{theorem}

\begin{proof}
 The group action is defined by $\sigma_\tau([a,b,c])=[\sigma_\tau(a), \sigma_\tau(b), \sigma_\tau(c)]$ for each term $[a,b,c]$ in (1,j,k).
It follows from Theorem \ref{thm5} that if any cubic MRS function $f(\vec{x})=(1,j,k)$ in standard form is affine equivalent to any cubic MRS function $g(\vec{x})=(1,p,q)$ by a permutation which preserves rotation symmetry, then there exists a permutation $\sigma_{\tau,n}$ in $G_n$ such that
$\sigma_{\tau,n}((1,j,k))=(1,p,q)$. Now the fact that the orbits are exactly the affine equivalence classes under permutations which preserve rotation symmetry follows from Lemmas \ref{lem4} and \ref{lem5}.
\end{proof}

\begin{remark}
\label{rem5}
We conjecture that if two cubic MRS functions in $B_n$ are affine equivalent, then there is a permutation of the $n$ variables which gives the equivalence.  This cubic analog of Theorem \ref{thm4} cannot be proved in the same way as the earlier result, since there seems to be no cubic version of Theorems \ref{thm2} and \ref{thm3}; thus proving the conjecture may be difficult.  It is well known that
the frequency count of the absolute values of the Walsh spectrum for a Boolean function is an affine invariant (see \cite[pp. 7 - 12]{CBF}).  Using this fact, we proved the conjecture for $n \leq 32$.   We did not need to do any computation for the cases $n \leq 21$, since the needed data for these cases are given in the Online Database of Boolean Functions \cite{ODBF}.  In all cases for $n \leq 32$, we verified that whenever two equivalence classes given by the action of  $G_n$ have the same weight and nonlinearity, then the classes have different frequency counts for the Walsh spectrum.  For some small values of $n$ this is very easy, since all equivalence classes given by the action of $G_n$ have different weights; this is true, for instance, for $n=9$ (see Table 1 below).
\end{remark}

\begin{remark}
\label{rem6}
 We want to determine the smallest group whose action \eqref{eqn15} gives the affine equivalence classes. In the trivial cases $n=3$ and $4$ there is only
one function in $C_n$ so we can take the smallest group to be the identity alone. In the case $n=5$ there are two functions $(1,2,3)$, $(1,2,4)$ in one 
equivalence class and the cyclic group generated by the 4-cycle permutation (2453) maps these functions to each other. For $n=6$ there are three classes:
$\{(1,2,3)\}, \{(1,3,5)\}$ and $\{(1,2,4), (1,2,5)\}$ and direct calculation shows that the group $G_6$ of order $2$ (generated by the product of three transpositions
$(16)(25)(34)$) is the smallest one which gives the equivalence classes. Similarly, for $n=7$ there are two classes $\{(1,2,3),(1,2,5),(1,3,5)\}$ and 
$\{(1,2,4),(1,2,6)\}$ and the cyclic group $G_7$ of order 6 (generated by the 6-cycle $\xi = (243756)$) is the smallest one which gives the equivalence classes. 
Finally, for $n=8$, there are four classes $\{(1,2,3),(1,3,6)\},\; \{(1,2,4),(1,2,7)\}$, $\{(1,2,5), (1,2,6)\}$ and $\{(1,3,5)\}$. 
The group $G_8=\{\sigma_1, \sigma_3, \sigma_5, \sigma_7\}$ is a noncyclic group of order $4$. Each of the nonidentity elements $\sigma_3, \sigma_5, \sigma_7$ of 
order 2 fixes both elements of exactly one of the three 2-element equivalence classes, so $G_8$ is the smallest group which gives the four equivalence classes.
\end{remark}
The next theorem shows that for $n\geq6$ the group $G_n$ is always the smallest one which gives the equivalence classes, by using the group action in Theorem \ref{thm7}.
Since $|G_n|=\varphi(n)$ ($\varphi$ is Euler's function) and the structure of the group $U_n$ in Theorem \ref{thm6} is well-known, this gives a detailed description of the affine 
equivalence classes of $C_n$ under permutations which preserve rotation symmetry.

\begin{theorem}
 For $n\geq6$, the group $G_n$ of order $\varphi(n)$ is the smallest group whose action \eqref{eqn15} gives the equivalence classes of $C_n$ under permutations which preserve rotation symmetry.
\end{theorem}

\begin{proof}
 We know from Theorem \ref{thm7} that  the orbits of the action \eqref{eqn15} of $G_n$ on $C_n$ are the affine equivalence classes, so we need only prove that no smaller group will give the equivalence classes. For $n=6,7,8$ this follows from the calculations
referred to in Remark \ref{rem6}. For $n>8$, we prove that the function $f(\vec{x})=(1,2,4)$ in n variables is always in an equivalence class of length $\varphi(n)$. Since
the order of $G_n$ is $\varphi(n)$, this shows that no smaller group can give the equivalence classes of $G_n$, as stated in the theorem.

We actually show that for $n>8$ the identity $e$ of  $G_n$ is the only element of $G_n$ which fixes $f=(1,2,4)$, that is, the stabilizer of $(1,2,4)$ is $e$. By elementary group theory this means the orbit of $(1,2,4)$ (which by Theorem \ref{thm7} is the same as its equivalence class) has length $\varphi(n)$, as required.

So we suppose that for some $\tau$ relatively prime to n we have

\begin{equation*}
 \sigma_\tau((1,2,4))=(1,\tau+1,3\tau+1)=(1,2,4).
\end{equation*}

This means that the term $[1,\tau+1,3\tau+1]$ satisfies

\begin{equation}
\label{eqn16}
 [1,\tau+1,3\tau+1]=[1,2,4]~~\text{or}~~[1,3,n]~~\text{or}~~[1,n-2,n-1]
\end{equation}

In the first case in \eqref{eqn16}, we must have either $\tau+1\equiv2\bmod{n}$ and $3\tau+1\equiv4\bmod{n}$ (so $\tau=1$ and $\sigma_1=e$)
or $\tau+1\equiv4\bmod{n}$ and $3\tau+1\equiv2\bmod{n}$ (so $\tau\equiv3\bmod{n}$ and $8\equiv0\bmod{n}$, giving $n=8$; in this case, $\sigma_3$ fixes $(1,2,4)$).

In the second case in \eqref{eqn16}, we must have either $\tau+1\equiv3\bmod{n}$ and $3\tau+1\equiv0\bmod{n}$ (so $\tau=2$ and $n=7$; in this case $\sigma_2$ fixes
$(1,2,4)$) or $\tau+1\equiv0\bmod{n}$ and 3$\tau+1\equiv3\bmod{n}$ (so $\tau\equiv-1\bmod{n}$   and $-2\equiv3\bmod{n}$, giving $n=5$ and $\tau=4$).

In the third case in \eqref{eqn16}, we must have either $\tau+1\equiv{n-2}\bmod{n}$ and $3\tau+1\equiv{n-1}\bmod{n}$ (so $\tau\equiv{-3}\bmod{n}$ and $7\equiv{0}\bmod{n}$,
giving $n=7$ and $\tau=4$; in this case $\sigma_4$ fixes $(1,2,4)$) or $\tau+1\equiv{n-1}\bmod{n}$ and $3\tau+1\equiv{n-2}\bmod{n}$ (so $\tau\equiv{-2}\bmod{n}$ and 
$3\equiv0\bmod{n}$, giving $n=3$). Thus if $n>8$, \eqref{eqn15} is only possible if $\sigma_\tau=e$ and the theorem is proved.
\end{proof}

\section{The equivalence classes for prime n}

If the number of variables is a prime p, then we can obtain a very detailed
description of the affine equivalence classes.  Define 
$$E(n) = number~of\;equivalence\;classes\;of\;cubic\;
 MRS\;functions\;in\;n\;variables.$$ 
We can evaluate $E(n)$ by using the well-known Burnside's Lemma applied to the group $G_n$ acting on $C_n$, as described in Theorem \ref{thm7}.
Let 
\begin{equation*}
Fix(\sigma) = number\;of\;functions\;in\;C_n\;fixed\;by\;\sigma. 
\end{equation*}

\begin{lemma}
\label{lemma41}
  We have 
\begin{equation*}
 E(n) = (1/|G_n|)\sum_{\sigma \in G_n}  Fix(\sigma). 
\end{equation*}

\end{lemma}

\begin{proof}
This is a special case of Burnside's Lemma for counting orbits.  By Theorem \ref{thm7}, the orbits in this special case are the affine equivalence classes. 
\end{proof}

The next theorem gives a complete description of the number and size of the affine equivalence classes when the number of variables is a prime.

\begin{theorem}
\label{thm8}
Suppose p is a prime.  Then 

\begin{equation*}
E(p) = [p/6] + 1. 
\end{equation*}
     Suppose $p > 5$.  There is exactly one equivalence class of size $(p-1)/2$, namely the class containing $(1,2,3)$.   
If $p \equiv 1 \mod{6}$, there is exactly one equivalence class of size $(p-1)/3$.  If $p \equiv 1 \mod{6}$, all the remaining 
$E(p) - 2$ equivalence classes have size $p - 1$. If $p \equiv 5 \mod{6}$, all the remaining $E(p) - 1$ equivalence classes have size $p - 1$.  
\end{theorem}

To prove Theorem \ref{thm8}, we will need the following three lemmas. 
          
\begin{lemma}
\label{lemma43}
  If $p$ is prime, then every affine equivalence class of cubic MRS functions in p variables contains a function $(1,2,m)$ for some $m > 2$.
\end{lemma}

\begin{proof}
It suffices to show that if $(1,j,k)$ is any function in $p$ variables, then there is some permutation $\sigma_{\tau} \in G_p$ such that 
\begin{equation}
\label{eqn17}
\sigma_{\tau}(\;(1,j,k)\;) = (1,2,m)                                                
\end{equation}
for some $m$.  Equation \eqref{eqn17} holds if there exists some $\tau$ not divisible by $p$ such that
\begin{equation}
\label{eqn18}
(1, (j-1)\tau +  1  \Mod p, (k-1)\tau + 1 \Mod p)  =  (1,2,m)
\end{equation}
for some integer $m > 2$.  Now \eqref{eqn18} implies $(j-1)\tau + 1 \equiv 2 \mod{p}$
and for any $j$ with $2 \leq j \leq p - 1$ this linear congruence has a unique solution
$\tau \Mod{p}$  with $\tau$ not divisible by $p$.  Given this solution $\tau$, an integer $m$ for which 
\eqref{eqn18} holds is  $m = (k-1)\tau + 1 \Mod{p}$, as given in \eqref{eqn18}.  Since
$\tau \not\equiv 0 \mod{p}$, we have $m \neq 1$.  Since $(j-1)\tau \equiv 1  \mod{p}$ and
$j \neq k$, we have $m \neq 2$. Thus $m > 2$, as required. 

\end{proof}

\begin{lemma}
\label{lemma44}
  For $n > 3$, the function $(1,2,3)$ is always in an affine equivalence class of size  $\varphi(n)/2$.
  The identity $e$ and $\sigma_{n-1}$ are the only elements of $G_n$ which fix the functions in this class.
\end{lemma}
 
\begin{proof} 

We have $\sigma_{n-1}(\;(1,2,3)\;) = (1,n,n-1) = (1,2,3)$, so  $\sigma_{n-1}$ fixes $(1,2,3)$.
Since $G_n$ is Abelian, this means $\sigma_{n-1}$ fixes all of the elements in the equivalence class of $(1,2,3)$.
Now suppose that for some $\tau$ relatively prime to $n$ we have 

\begin{equation*}
\sigma_\tau(\;(1,2,3)\;) = (1, \tau + 1, 2\tau + 1) = (1,2,3). 
\end{equation*}
This means that the term $[1, \tau + 1, 2\tau + 1]$ satisfies 
\begin{equation}
\label{eqn19}
[1, \tau + 1, 2\tau + 1] = [1,2,3]~~\text{or}~~[1,2,n]~~\text{or}~~[1,n-1,n].
\end{equation}

 In the first case in \eqref{eqn19} we have either $\tau + 1 \equiv 2 \mod{n}$ and
$2\tau + 1 \equiv 3 \mod{n}$ (so $\tau = 1$ and $\sigma_1 = e$) or $\tau + 1 \equiv 3\mod{n}$ 
and $2\tau + 1 \equiv 2 \mod{n}$ (so $\tau = 2$ and $3 \equiv 0 \mod{n}$, which gives $n = 3$).
In the second case in \eqref{eqn19} we have either  $\tau + 1 \equiv 2 \mod{n}$ and
$2\tau + 1 \equiv 0 \mod{n}$  (so $\tau = 1$ and $3 \equiv 0 \mod{n}$, which gives $n = 3$) or
$\tau + 1 \equiv 0 \mod{n}$ and $2\tau + 1 \equiv 2\mod{n}$ (so $\tau = n - 1$ and $3 \equiv 0 \mod{n}$, which gives $n = 3$).  
In the third case in \eqref{eqn19} we have either $\tau + 1 \equiv n-1 \mod{n}$ and $2\tau + 1 \equiv 0\mod{n}$ 
(so $\tau = n-2$ and $3 \equiv 0 \mod{n}$, which gives $n = 3$) or $\tau + 1 \equiv 0 \mod{n}$  and $2\tau + 1 \equiv n-1\mod{n}$
(so $\tau = n - 1$;  we already saw that  $\sigma_{n-1}$ fixes $(1,2,3)$ ).

Thus there are exactly two elements of $G_n$ which fix $(1,2,3)$, that is, the
\emph{stabilizer} of $(1,2,3)$ has order $2$.  Since $|G_n| = \varphi(n)$, by elementary group theory the \emph{orbit} of $(1,2,3)$
(which is the same as its equivalence class) has length $\varphi(n)/2$.
\end{proof}

\begin{lemma}
\label{lemma45}
Assume $p \equiv 1 \mod{6}$ is prime. Then the order of $\sigma_\tau$  in $G_p$ is $3$ if and only if  $\tau^3 \equiv 1\mod{p}$.
There are exactly two such elements of order $3$ in $G_p$ and they have the form $\sigma_k$ and $\sigma_{k^2 \Mod{p}}$ for an integer
$k > 1$ which satisfies $k^3 \equiv 1\mod{p}$.  Both of these permutations fix the functions in the same equivalence class of size $(p-1)/3$, namely the class containing $(1,2,k+2)$.
\end{lemma}

\begin{proof}
Since $G_p$ is cyclic by Theorem \ref{thm6}, elements of order $3$ exist if and only if
$p \equiv 1\mod{6}$, and then there are exactly two elements $\sigma_\tau \in G_p$ with order $3$.
If $\sigma_k$ has order $3$, then $\sigma_k(2) = k+1 \Mod{p}$,
$\sigma_k^2(2) = k^2 + 1 \Mod{p}$ and $\sigma_k^3(2) = (k^3 + 1) \Mod{p} = 2$.  Thus
$k^3 \equiv 1\mod{p}$ and the other element of order $3$ is $\sigma_{k^2\Mod{p}}$. Since
$k^3 - 1 = (k - 1)(k^2 + k + 1)$ we have $k^2 + k + 1 \equiv 0\mod{p}$.  Therefore
$\sigma_k(\;(1,2,k+2)\;) = (1, k + 1, k^2 + k + 1) = (1, k+1, p) = (1,2,k+2)$ and
 $\sigma_{k^2 \Mod{p}} = \sigma_k(\sigma_k)$ also fixes $(1,2,k+2)$.
Thus  the two elements  of order $3$ fix all the functions in the class containing $(1,2,k+2)$. Hence 
the stabilizer of $(1,2,k+2)$ has order 3 and therefore the class containing $(1,2,k+2)$ has size 
$(p-1)/3$.

Conversely, if $\tau^3 \equiv 1\mod{p}$, then reversing the above argument shows that
$\sigma_\tau$ and $\sigma_{\tau^2}$ have order $3$.  Note that  $\tau^3 - 1 = (\tau - 1)(\tau^2 + \tau +1) \equiv 0\mod{p}$ 
gives the congruence $4(\tau^2 + \tau +1) = (2\tau +1)^2 + 3 \equiv 0\mod{p}$, which always has two roots
when $p \equiv 1\mod{6}$ since then $-3$ is a quadratic residue mod p  by quadratic reciprocity.
\end{proof} 

\noindent \emph{Proof of Theorem \ref{thm8}.}
We use Lemma \ref{lemma41} with $n = p$ prime to evaluate $E(p)$.
Thus it suffices to determine $Fix(\sigma)$ for all $\sigma \in G_p$.  By Lemma \ref{lemma43},
if $\sigma \in G_p$ fixes any equivalence class containing a function $(1,j,k) \in D_n$,
we may assume that $\sigma$ fixes some function $(1,2,m)$ with $m>2$.  So suppose 

\begin{equation*}
          \sigma_\tau(\;(1,2,m)\;) = (1, \tau + 1, (m-1)\tau + 1) = (1,2,m) 
\end{equation*}
for some $\tau$  with $1 \leq \tau < p$.  This means that the term $[1, \tau + 1, (m-1)\tau+ 1]$ satisfies 
\begin{equation}
 \label{eqn20}
[1, \tau + 1, (m-1)\tau+ 1] =[1,2,m]~~\text{or}~~[1,m-1,p]~~\text{or}~~[1,p-m+2,p-m+3].
\end{equation}

In the first case in \eqref{eqn20}, we have either $\tau + 1 \equiv 2 \mod{p}$ and
$(m-1)\tau + 1 \equiv m \mod{p}$ (so $\tau = 1$ and $\sigma_1 = e$) or  $\tau + 1 \equiv m \mod{p}$ and $(m-1)\tau + 1 \equiv 2 \mod{p}$ (so $\tau \equiv m - 1 \mod{p}$ and
$(m-1)^2 \equiv 1 \mod{p}$;  since p is prime this gives either $m=p$ [so $\sigma_\tau$ fixes $(1,2,p) = (1,2,3)$, which by Lemma \ref{lemma44} gives $\sigma_\tau$ equal to $e$ or $\sigma_{p-1}$]
or $m = 2$ [impossible]).

In the second case in \eqref{eqn20}, we have either $\tau + 1 \equiv m - 1 \mod{p}$ and
$(m-1)\tau + 1 \equiv p \mod{p}$ or $\tau + 1 \equiv p \mod{p}$ and $(m-1)\tau + 1 \equiv m-1 \mod{p}$.  The first pair of congruences gives $\tau \equiv m - 2 \mod{p}$ and
$(m-1)(m-2) \equiv -1 \mod{p}$.  This implies
\begin{equation}
\label{eqn21}
m^2 - 3m + 3 \equiv 0 \mod{p},
\end{equation}
so $(2m - 3)^2 \equiv -3 \mod{p}$ and there are exactly two values of $m$ which give solutions; therefore $-3$ is a quadratic residue $\mod{p}$ and so by
quadratic reciprocity we have $p \equiv 1 \mod{6}$.  Also \eqref{eqn21} implies  
\begin{equation*}
\tau^3 \equiv (m-2)^3 \equiv -3m^2 +9m -8 \equiv 1 \mod{p},  
\end{equation*}
so by Lemma \ref{lemma45} $\sigma_\tau$ has order $3$.  The second pair of congruences gives
$\tau = p - 1$ and $2m \equiv 3 \mod{p}$, that is, $m = (p + 3)/2$.  This implies that
$\sigma_{p-1}$ fixes the class containing $(1,2,(p+3)/2)$, but this is the same as the class 
containing $(1,2,3)$, since $\sigma_{(p+1)/2}( (1,2,3) ) = (1,2,(p+3)/2)$.

In the third case in \eqref{eqn20}, we have either $\tau + 1 \equiv p - m + 2 \mod{p}$ and
$(m-1)\tau + 1 \equiv p - m + 3 \mod{p}$ or $\tau + 1 \equiv p - m + 3 \mod{p}$ and
$(m-1)\tau + 1 \equiv p - m + 2 \mod{p}$. The first pair of congruences gives
$\tau  \equiv p - m + 1 \mod{p}$ and $(m - 1)(p - m + 1) + 1 \equiv p - m + 3 \mod{p}$. This gives \eqref{eqn21} again,
so $p \equiv 1 \mod{6}$ and $\sigma_\tau$ again is one of the two elements of order $3$ in $G_p$.  
The second pair of congruences gives $\tau \equiv p - m + 2 \mod{p}$
and $(m - 1)(p - m + 2) + 1 \equiv p - m + 2 \mod{p}$.  This gives
$(m-1)(m-3) \equiv 0 \mod{p}$, so either $m = 1$ (impossible) or $m = 3$.  Thus 
$\sigma_\tau$ fixes $(1,2,3)$, which by Lemma \ref{lemma44} gives  $\sigma_\tau$ equal to $e$ or $\sigma_{p-1}$.

Combining the results above, we see that $Fix(\sigma) = 0$ unless $\sigma$ is
$e$ or $\sigma_{p-1}$ (for any prime $p>5$) or one of the two elements of order $3$, namely $\sigma_k$
and $\sigma_{k^2 \Mod{p}}$ with $k^3 \equiv 1 \mod{p}$ (for $p \equiv 1 \mod{6}$).
We also have   
\begin{equation}
\label{eqn22}
Fix(e) = |D_p| = (p^2 - 3p + 2)/6
\end{equation}
by Lemma \ref{lem3}, $Fix(\sigma_{p-1}) = (p-1)/2$ by Lemma \ref{lemma44}
(since we proved above that the only class fixed by $\sigma_{p-1}$ contains $(1,2,3)$) and 
$Fix(\sigma_k) = Fix(\sigma_{k^2 \Mod{p}}) = (p - 1)/3$ by Lemma \ref{lemma45}.  
Plugging our data into Lemma \ref{lemma41} gives $E(p) = [p/6] + 1$, and the other assertions in the theorem then follow by computation using \eqref{eqn22}. 

\section{Equivalence Classes for $n=3^{k}$}

For the convenience of the reader, we give tables of the equivalence classes for $n = 9$ and $27$ below.  These may be helpful in following the proofs in this section.
The classes are arranged in lexicographical order of their representatives in $D_n$.  My research assistants Max Bileschi and Dan Padgett computed tables like these for hundreds of values of 
$n$, and these were of great value in formulating and checking several of the theorems in this paper.

\begin{table}[p] 
\caption{Affine equivalence classes for cubic MRS functions in 9 variables} 
\centering  							
\begin{tabular}{c rrrrrr}  					
\hline\hline                   			 		
Class & \multicolumn{6}{c}{Functions} \\ [0.5ex]    
\hline    							
Class 1, size 3  & (1,2,3) & (1,2,6) &  (1,3,5)\\	  	
Class 2, size 6  & (1,2,4) & (1,2,5) &  (1,2,7) & (1,2,8) & (1,3,6) & (1,3,7)\\
Class 3, size 1  & (1,4,7)\\[1ex] 				
\hline                          				
\end{tabular} 
\label{tab:hresult} 
\end{table}

\begin{table}[bp]
\caption{Affine equivalence classes for cubic MRS functions in 27 variables} 
\centering    		                        
\begin{tabular}{@{\extracolsep{\fill}} l c c c rrrrr}       	      
\hline\hline                                
 Class & Size & \multicolumn{6}{c}{Functions} \\ [0.5ex]
\hline                                      
& & (1,2,3) & (1,2,15) & (1,3,5) & (1,5,9) & (1,6,11)  \\[-1ex]
\raisebox{1.5ex}{Class 1} & \raisebox{1.5ex}{9}
& (1,6,17) & (1,8,15) & (1,8,18) & (1,9,17) \\[1ex] 

& &(1,2,4) & (1,2,14) & (1,2,16) & (1,2,26) & (1,3,7) & (1,3,24)  \\
Class 2 & 18 & (1,4,11) & (1,4,12) & (1,4,20) & (1,4,21) & (1,5,13) & (1,5,20)\\
& & (1,6,12) & (1,6,16) & (1,6,18) & (1,6,22) & (1,7,14) & (1,7,21)\\[1ex]

& &(1,2,5) & (1,2,8) & (1,2,22) & (1,2,25) & (1,3,9) & (1,3,15)  \\
Class 3 & 18 & (1,3,16) & (1,3,22) & (1,4,9) & (1,4,14) & (1,4,18) & (1,4,23)\\
& & (1,5,16) & (1,5,17) & (1,6,13) & (1,6,21) & (1,7,17) & (1,7,18)\\[1ex]

& &(1,2,6) & (1,2,9) & (1,2,12) & (1,2,18) & (1,2,21) & (1,2,24)  \\
Class 4 & 18 & (1,3,8) & (1,3,11) & (1,3,14) & (1,3,17) & (1,3,20) & (1,3,23)\\
& & (1,5,12) & (1,5,15) & (1,5,18) & (1,5,21) & (1,6,14) & (1,6,20)\\[1ex]

& &(1,2,7) & (1,2,13) & (1,2,17) & (1,2,23) & (1,3,6) & (1,3,13)  \\
Class 5 & 18 & (1,3,18) & (1,3,25) & (1,4,8) & (1,4,15) & (1,4,17) & (1,4,24)\\
& & (1,5,11) & (1,5,22) & (1,7,15) & (1,7,20) & (1,8,16) & (1,8,20)\\[1ex]

& &(1,2,10) & (1,2,11) & (1,2,19) & (1,2,20) & (1,3,10) & (1,3,12)  \\
Class 6 & 18 &  (1,3,19) & (1,3,21) & (1,5,10) & (1,5,14) & (1,5,19) & (1,5,23)\\
& & (1,6,15) & (1,6,19) & (1,8,17) & (1,8,19) & (1,9,18) & (1,9,19)\\[1ex]

Class 7 & 3 & (1,4,7) & (1,4,16) & (1,7,13)\\[1ex]

Class 8 & 6 & (1,4,10) & (1,4,13) & (1,4,19) & (1,4,22) & (1,7,16) & (1,7,19)\\[1ex]

Class 9 & 1 & (1,10,19) \\[1ex]
  
\hline                          
\end{tabular}
\label{tab:PPer} 
\\[.2in]

A Mathematica program which very quickly calculates tables like these for $n$ not too large (say
$n < 800$) is available from the author.

\end{table}

If the number of variables is a power of $3$, then we can obtain a very detailed 
description of the equivalence classes.  We begin with a few preliminary lemmas.
We assume throughout this section that $n$ denotes the number of variables for our cubic MRS functions. 

\begin{lemma}
 \label{lemma51}
If $n = 3^k$ for $k \geq 1$, then there is a unique equivalence class of size $1$, and it has representative
$f_1 = (1,\;3^{k-1} + 1,\;2\cdot 3^{k-1}+1)$ in $D_n$.  This is the short function in n variables.
\end{lemma}

\begin{proof}
It suffices to show that $f_1$ is the only function fixed by all $\sigma \in G_n$.
It is easy to see that every $\sigma$ fixes $f_1$, so it suffices to show that $\sigma_2$
defined as in Section 3 by
\begin{equation*}
 \sigma_2(i) = 2(i - 1) + 1 \Mod{n}~~\text{for}~~i = 1, 2, . . . , n 
\end{equation*}
fixes only $f_1$.  If $\sigma_2$ fixes $(1,r,s)$ with $r < s$ we have 
\begin{align}
\label{eqn23}
 [1,r,s] = &[1,2r - 1, 2s - 1]~~\text{or}~~[1, 2(s - r)+1,n-2r+3]~~\text{or}~~ \\
\notag &[1,n-2s+3,n-2(s-r)+1].
\end{align}

  In the first case in \eqref{eqn23} we have either $r \equiv 2r - 1 \mod{n}$ and $s \equiv 2s - 1 \mod{n}$ 
(so $r \equiv s \equiv 1 \mod{n}$, which is impossible) or $r \equiv 2s - 1 \mod{n}$
and $s \equiv 2r - 1 \mod{n}$ (this implies  $r \equiv  s \equiv 1 \mod{3^{k-1}}$ and so,
since $r < s$, we must have $r =  3^{k-1} + 1$ and $s = 2\cdot 3^{k-1}+1$; this gives $f_1$).

  In the second case in \eqref{eqn23} we have either r $\equiv 2s-2r+1 \mod{n}$ and $s \equiv -2r+3 \mod{n}$ 
(this implies $7r \equiv 7 \mod{n}$, so $r \equiv 1 \mod{n}$, which is impossible) or
$r \equiv -2r + 3 \mod{n}$ and $s \equiv 2s-2r+1 \mod{n}$ (this implies $r \equiv  s \equiv 1 \mod{3^{k-1}}$,
 so as before we get $f_1$).  

  In the third case in \eqref{eqn23}, we have either
$r \equiv -2s+3 \mod{n}$ and $s \equiv -2s+2r+1 \mod{n}$ (this implies $7s \equiv 7 \mod{n}$,
 so $s \equiv 1 \mod{n}$, which is impossible) or $r \equiv -2s+2r+1 \mod{n}$ and $s \equiv
-2s+3 \mod{n}$ (this implies $s \equiv 1 \mod{3^{k-1}}$, which again gives $f_1$).   
\end{proof}

\begin{lemma}
 \label{lemma52}
If $n = 3^k$ for $k \geq 1$, then there is no equivalence class of size $2$.
\end{lemma}

\begin{proof}
It suffices to show that the only function $f$ such that $\sigma_\tau^2(f) = f$ for all $\tau$
not divisible by $3$ is $f = f_1$ (the short function in n variables); then by Lemma \ref{lemma51} the size of the equivalence class
of $f$ is $1$.

So we suppose that for some function $(1,r,s)$ with $r < s$ we have
\begin{equation*}
\sigma_\tau^2(\;(1,r,s)\;)\;=\;(1,(r-1)\tau^2 + 1, (s-1)\tau^2 + 1)\;=\;(1,r,s). 
\end{equation*}
We take $\tau = 2$, and then it follows that
\begin{equation*}
[1, 4r-3, 4s-3]\;=\;[1, r, s]~~\text{or}~~[1, s-r+1, 3^k-r+2]~~\text{or}~~[1, 3^k-s+2, 3^k+r-s+1]. 
\end{equation*}
As in the proof of Lemma \ref{lemma51}, this gives three alternatives, each involving two choices of a pair of congruences, and we call these six cases A1, A2, B1, B2, C1 and C2, respectively.

Case A1 gives $4r-3 \equiv r \mod{3^k}$  and $4s-3 \equiv s \mod{3^k}$, so $r \equiv s \equiv 1 \mod{3^{k-1}}$.  This means $r = 3^{k-1}+1$ and $s = 2 \cdot 3^{k-1}+1$ (since we know $r < s$),
so $(1,r,s) = f_1$.
 Case A2 gives $4r-3 \equiv s \mod{3^k}$  and $4s-3 \equiv r \mod{3^k}$, so 
$4(4s-3) \equiv s + 3 \mod{3^k}$, which implies $s \equiv 1 \mod{3^{k-1}}$.  By a symmetric argument, we also get $r \equiv 1 \mod{3^{k-1}}$.  By the argument in Case A2, we again obtain
$(1,r,s) = f_1$.

Case B1 gives $4r-3 \equiv s-r+1 \mod{3^k}$  and $4s-3 \equiv -r+2 \mod{3^k}$, so
$s \equiv 5r-4 \mod{3^k}$ and $4(5r-4) \equiv -r +5 \mod{3^k}$.  This implies $r \equiv 1 \mod{3^{k-1}}$; 
substituting this into $4s-3 \equiv -r+2 \mod{3^k}$ gives $s \equiv 1 \mod{3^{k-1}}$.
Now $(1,r,s) = f_1$ as in Case A2.  Case B2 gives $4r-3 \equiv -r+2 \mod{3^k}$  and $4s-3 \equiv s-r+1 \mod{3^k}$; 
the former congruence implies $r \equiv 1  \mod{3^k}$, which is impossible.

Case C1 is the same as Case B2 with $r$ and $s$ interchanged, so we get $s \equiv 1  \mod{3^k}$, which is impossible.
Case C2 is the same as Case B1 with $r$ and $s$ interchanged, so again we get $(1,r,s) = f_1$.
\end{proof}

\begin{lemma}
 \label{lemma53}
If $n = 3^k$ and $k \geq j \geq 2$, then the equivalence class
of $f_j = (1, 3^{k-j} + 1, 2 \cdot 3^{k-j} + 1)$ has size $3^{j-1}$.
This is the only class of size $3^{j-1}$ for $n = 3^k$.
\end{lemma}

\begin{proof}
We first show that $\sigma_{\tau}^{3^{j-1}}(f_j) = f_j$ for any $\tau$ (of course we have $gcd(\tau, 3) = 1$
since $\sigma_{\tau}$ is in $G_n$).  We have
 \begin{equation}
\label{eqn24} 
\sigma_{\tau}^{3^{j-1}}(f_j) = (1, 3^{k-j}\tau^{3^{j-1}} + 1, 2 \cdot 3^{k-j}\tau^{3^{j-1}} + 1)
\end{equation}
and  $\tau^{3^{j-1}} \equiv  (\tau/ 3)  \mod{3^j}$  (here $(\tau/ 3)$ is a Legendre symbol).  So the right-hand side of \eqref{eqn24} is $f_j$ if  $\tau \equiv 1 \mod{3}$ or is
\begin{equation*}       
(1, 3^k - 3^{k-j} + 1, 3^k - 2 \cdot 3^{k-j} + 1) = f_j
\end{equation*}
if $\tau \equiv 2 \mod{3}$.

This shows that the size of the equivalence class of $f_j$ divides $3^{j-1}$.  If $j = 2$, the size must be $3$
for any $k \geq 2$ because from Lemma \ref{lemma51} we know that the only class of size $1$ contains the short function.
We know from Table 1 above that for $n = 9$ the class of $f_2 = (1, 2, 3)$ is the only class of size $3$. 
We now show that also for $k \geq 3$, the class of $f_2$ is the only class of size $3$. 

We suppose that
\begin{equation}
\label{eqn25}
\sigma_{\tau}^3(\;(1,r,s)\;) = (1, (r-1) \tau^3+1, (s-1) \tau^3+1) = (1,r,s) 
\end{equation}
for every $\tau$, and we prove that $(1,r,s)$ is either equivalent to the short function or $(1,r,s)$ belongs to the 
class of $f_2$, that is, $(1,r,s)$ must be affine equivalent to $f_2$.  We do this by induction, so we assume that the
result is true when $n = 3^{k-1}$ and we prove it for $n = 3^k$. We already have the base case when $k = 2$.

It follows from $\eqref{eqn25}$ that
\begin{align}
\label{eqn26}
&[1,(r-1) \tau^3+1,(s-1) \tau^3+1] = [1,r,s]~~\text{or}~~\\
\notag&[1,s-r+1,3^k-r+2]~~\text{or}~~[1,3^k+r-s+1, 3^k-s+2].
\end{align}

As in the proof of Lemma \ref{lemma52}, this gives six possibilities, which we call cases A1, A2, B1, B2, C1 and C2. 

First we take $\tau = n-2$ in \eqref{eqn26}. Then Case A1 gives $-8r + 9 \equiv r \mod{3^k}$ and
$-8s + 9 \equiv s \mod{3^k}$, so $r \equiv s \equiv 1 \mod{3^{k-2}}$.  This implies 
\begin{equation}
\label{eqn27}
r = a \cdot 3^{k-2}+1~~\text{and}~~s = b \cdot 3^{k-2}+1~~\text{for some a and b},
\end{equation}
which includes $(1,r,s) = f_2$.  Case A2 gives $-8r + 9 \equiv s \mod{3^k}$ and $-8s + 9 \equiv r \mod{3^k}$. 
Substituting the first congruence into the second gives $r \equiv 1 \mod 3^{k-2}$,
and by symmetry also $s \equiv 1 \mod{3^{k-2}}$.  Therefore \eqref{eqn27} is again true and
substituting those formulas into the first congruence gives $-8a \equiv a \equiv b \mod{9}$.
This is impossible since then we would have $r \equiv s \mod{3^k}$ from \eqref{eqn27}. 

Case B1 gives $-8r + 9 \equiv s-r+1 \mod{3^k}$ and $-8s + 9 \equiv -r+2 \mod{3^k}$, so
$s \equiv -7r +8 \mod{3^k}$.  Substituting this into the first congruence gives $r \equiv 1 \mod 3^{k-1}$
and so also s $\equiv 1 \mod 3^{k-1}$.  This implies that ${r, s}$ must be ${3^{k-1}+1, 2 \cdot 3^{k-1}+1}$,
so $(1,r,s)$ is the short function, which by Lemma \ref{lemma51} has class size $1$.  Case B2 gives $-8r + 9 \equiv -r+2 \mod 3^k$
and $-8s + 9 \equiv s-r+1 \mod{3^k}$. The first congruence already gives a contradiction since it implies $r \equiv 1 \mod 3^k$,
which is impossible. 

Case C1 gives second congruence $-8s+9 \equiv -s+2 \mod 3^k$, which gives the contradiction $s \equiv 1 \mod 3^k$,
as in Case B2. Case C2 gives $-8r + 9 \equiv -s+2 \mod 3^k$ and $-8s + 9 \equiv r-s+1 \mod 3^k$, so we have Case B1 
with $r$ and $s$ switched, which gives \eqref{eqn27} with $r$ and $s$ switched. 

Now without loss of generality we can assume that if $(1,r,s)$ satisfies \eqref{eqn25}, then \eqref{eqn27} holds. 
We now take $\tau = 2$ in \eqref{eqn26}. Then Case A1 gives $8r-7 \equiv r \mod{3^k}$ and
$8s-7 \equiv s \mod{3^k}$. This implies $r \equiv s \equiv 1 \mod{3^k}$, which is impossible. Case A2 gives 
$8r-7 \equiv s \mod{3^k}$ and $8s-7 \equiv r \mod{3^k}$. Substituting the second congruence into the first gives
$63s \equiv 63 \mod{3^k}$, which leads to \eqref{eqn27} again. If we assume \eqref{eqn27} and substitute into
the first congruence, we obtain $a + b \equiv 0 \mod{9}$.  Thus we can take $b = 9 - a$ and we get 
$(1, r, s) = (1, a \cdot 3^{k-2}+1, 3^k - a \cdot 3^{k-2}+1) = (1,  a \cdot 3^{k-2}+1, 2a \cdot 3^{k-2}+1)$.  Hence we must have 
\begin{equation}
\label{eqn28}
b \equiv 2a \mod{9}~~\text{and}~~a \not\equiv 0 \mod{9}.
\end{equation}

Case B1 gives $8r - 7 \equiv s-r+1 \mod{3^k}$ and $8s - 7 \equiv -r+2 \mod{3^k}$, so
$s \equiv 9r-8 \mod{3^k}$. Substituting this into the first congruence gives $r \equiv 1 \mod{3^k}$, which is impossible.
Case B2 gives $8r - 7 \equiv -r+2 \mod{3^k}$ and $8s - 7 \equiv s-r+1 \mod{3^k}$.  The first congruence implies 
$r \equiv 1 \mod{3^{k-2}}$ and the second implies $7(s - 1) \equiv -(r - 1) \mod 3^k$; these two congruences together give
$s \equiv 1 \mod{3^{k-2}}$, so using \eqref{eqn27} the second of these new congruences implies 
$7b + a \equiv 0 \mod{9}$ and $a \not\equiv 0 \mod{9}$, that is we get \eqref{eqn28} again with $a$ and $b$ interchanged.
Note the case $a = 6$, $b = 3$ gives the short function $f_1 = (1,r,s)$, and the case $a = 2$, $b = 1$ gives $f_2 = (1,r,s)$. 

Case C1 is identical to Case B2 with $r$ and $s$ interchanged,  so we can conclude that
in this case $b \equiv 2a\mod{9}$.  Case C2 is identical to Case B1 with $r$ and $s$ interchanged, so this case is impossible. 

Combining the above results, we see that without loss of generality we can assume that if $(1,r,s)$ satisfies \eqref{eqn25},
then \eqref{eqn27} and \eqref{eqn28} hold.  Thus we have eight possibilities for $(1,r,s)$, say
$g_a = (1, a \cdot 3^{k-1} + 1, 2a \cdot  3^{k-1} + 1),~~1 \leq a \leq 8$.  If $a = 3~~\text{or}~~6$, we have the short function.
For the other values of $a$, $g_a$ is affine equivalent to $f_2$.  To prove this, we simply choose $\tau$ in \eqref{eqn25}
such that $a \tau \equiv 1 \mod{9}$ (possible since $3$ does not divide $a$) and then 
\begin{equation*}
\sigma_\tau(g_a) = (1, a \tau \cdot 3^{k-2} + 1,  2a \tau \cdot 3^{k-2} + 1) = f_2. 
\end{equation*}
This completes the proof that for $k \geq 3$, the class of $f_2$ is the only class of size $3$.

Next we prove by induction that the equivalence class of $f_j$ is the only class of size $3^{j-1}$ for each $j > 2$,
assuming that we have already proved that the equivalence class of $f_t$ is the only class of size $3^{t-1}$ for each $t < j$.
The six cases above give the base case $j = 2$.  We simply repeat, \emph{mutatis mutandis}, 
the calculations above for the case $j = 2$.

Since the notation becomes complicated and the induction is straightforward, we only sketch the details.   

As before, first we take $\tau = n-2$ in \eqref{eqn26}. Case A1 leads to the general version of \eqref{eqn27}, namely 
\begin{equation}
\label{eqn29} 
r = a \cdot 3^{k-j}+1~~\text{and}~~s = b \cdot 3^{k-j}+1~~\text{for some $a$ and $b$}.
\end{equation}
To see that Case A2 is impossible, we need the fact that $2^{3^{j-1}} + 1 \equiv 0 \mod{3^j}$.
To see that Case B1 gives the short function, we need the fact that $3^{j+1}$ is the exact power of $3$ which 
divides $2^{3^j} + 1$. Case B2 turns out to be impossible as before, and Cases C1 and C2 reduce to 
Cases B2 and B1, respectively.

Next we assume \eqref{eqn29} and take $\tau = 2$. Case A1 is impossible and Case A2 leads to \eqref{eqn29} again.
From this we obtain  $a + b \equiv 0 \mod 3^j$  and that leads to the general version of \eqref{eqn28}, namely   
\begin{equation}
\label{eqn30}
b \equiv 2a \mod{3^j}~~\text{and}~~a \not\equiv 0\mod{3^j},
\end{equation}
where again we use the fact that $3^{j+1}$ is the exact power of $3$ which divides $2^{3^j} + 1$.

Case B1 is impossible and as above Case B2 leads to \eqref{eqn30} again with $a$ and $b$ interchanged.
As before, the cases C1 and C2 reduce to the cases B2 and B1, respectively, with $r$ and $s$ interchanged,
so we can assume without loss of generality that \eqref{eqn29} and \eqref{eqn30} hold.

Hence we have $3^j - 1$ functions, say $h_a = (1, a \cdot 3^{k-j} + 1, 2a \cdot  3^{k-j} + 1)$,
$1 \leq a \leq 3^j - 1$.  We use the notation  $3^m || a$ to mean that $3^m$ is the exact power of $3$ which divides $a$.
If $a = 3^{j-1}$ or $2 \cdot 3^{j-1}$ (that is, $3^{j-1} || a$), then $h_a$ is the short function.  If $3^{j-2} || a$,
then $h_a$ is of form $(1, c \cdot 3^{k-2}+1, 2c \cdot  3^{k-2}+1)$ for
some $c$ not divisible by $3$, and by our induction hypothesis $h_a$ is affine equivalent to $f_2$ and so is in the unique 
class of size $3$.  Similarly, if $3^{j-t} || a$ for $1 \leq t \leq j-1$, then $h_a$ is of form  
$(1, c \cdot 3^{k-t}+1, 2c \cdot  3^{k-t}+1)$ for some $c$ not divisible by $3$, and by our induction hypothesis $h_a$
is affine equivalent to $f_t$  and so is in the unique class of size $3^{t-1}$.  Finally, if $3$ does not divide $a$,
then $h_a$ is affine equivalent to $f_j$.  To prove this, we simply choose $\tau$ in \eqref{eqn25} such that  
$a\tau \equiv 1 \mod 3^j$ (possible since $3$ does not divide $a$) and then $\sigma_\tau(h_a) = f_j$.
This completes the proof that the class of $f_j$ is the only class of size $3^{j-1}$. 
\end{proof}

\begin{lemma}
 \label{lemma54}
Suppose $n = 3^k$ and $f = (1, r ,s)$ is in an equivalence class of size $2 \cdot 3^j=v(j)$, say, for $1 \leq j \leq k - 2$.  
Then for every $\tau$ not divisible by $3$ we have
\begin{equation}
 \label{eqn31}
   (r - 1) \tau^{v(j)} + 1 \equiv r \mod 3^k~~\text{and}~~(s - 1) \tau^{v(j)} + 1 \equiv s \mod 3^k
\end{equation}
and 
\begin{equation}
 \label{eqn32}
r \equiv s \equiv 1 \mod 3^{k-j-1}.
\end{equation}
\end{lemma}

\begin{proof}
Our hypotheses imply that for every $\tau$ not divisible by $3$ we have
\begin{equation}
\label{eqn33}
\sigma_\tau^v(1,r,s) = (1, (r-1) \tau^v + 1, (s-1) \tau^v + 1) = (1, r, s)
\end{equation}
with $v = v(j) = 2 \cdot 3^j$ and for at least one $\tau$ \eqref{eqn33} is not true for any $v$ such that
$0 < v < v(j)$.  As in the proof of Lemma \ref{lemma52}, \eqref{eqn33} gives six cases  A1, A2, B1, B2, C1 and C2.
Equation \eqref{eqn31} simply says that Case A1 applies, so we shall show \eqref{eqn32} is true in Case A1 and
that none of the other cases are possible.

In Case A1 the first congruence in \eqref{eqn31} gives  $(\tau^{v(j)} - 1)r \equiv \tau^{v(j)} - 1 \mod{3^k}$,
which implies $r \equiv 1  \mod 3^{k-j-1}$ since
\begin{equation} 
\label{eqn34}
\tau^{v(j)} \equiv 1 \mod 3^{j+1}
\end{equation}
by Euler's theorem.  Similarly, the second congruence in \eqref{eqn31} gives $s \equiv 1 \mod{3^{k-j-1}}$,
so \eqref{eqn32} holds.

Case A2 gives $\tau^{v(j)}r - (\tau^{v(j)} - 1) \equiv s \mod{3^k}$ and
$\tau^{v(j)}s - (\tau^{v(j)} - 1) \equiv r \mod{3^k}$.  Substituting the second congruence into the
first gives   $(\tau^{2v(j)} - 1)s \equiv  (\tau^{2v(j)} - 1) \mod{3^k}$, which implies
$s \equiv 1 \mod{3^{k-j-1}}$ by \eqref{eqn34}. By symmetry also $r \equiv 1 \mod{3^{k-j-1}}$, so if we put
$r = a \cdot 3^{k-j-1} + 1$ and  $s =  b \cdot 3^{k-j-1} + 1$, say, then the first congruence
implies  $a \equiv b \mod{3^{j+1}}$.  This gives $r \equiv s \mod{3^k}$, which is impossible.

Case B1 gives  $\tau^{v(j)}r - (\tau^{v(j)} - 1) \equiv s-r+1 \mod{3^k}$ and $\tau^{v(j)}s - (\tau^{v(j)} - 1) \equiv -r+2\mod{3^k}$.
The first congruence gives
\begin{equation}  
\label{eqn35}
s \equiv (\tau^{v(j)} + 1)r - \tau^{v(j)} \mod 3^k,
\end{equation}
and substituting this into the second congruence gives 
\begin{equation}
 \label{eqn36}
(\tau^{2v(j)}+\tau^{v(j)}+1)r \equiv \tau^{2v(j)}+\tau^{v(j)}+1 \mod{3^k}.
\end{equation}

\noindent Since $3$ does not divide $\tau$ and $v(j)$ is even,  $3$ but not $9$ divides $\tau^{2v(j)}+\tau^{v(j)}+1$, so
\eqref{eqn36} gives $r \equiv 1 \mod{3^{k-1}}$.  Now \eqref{eqn35} gives also $s \equiv 1 \mod{3^{k-1}}$, so
$(1,r,s)$ must be the short function, which contradicts our hypothesis.

Case B2 gives $\tau^{v(j)}r - (\tau^{v(j)} - 1) \equiv -r+2 \mod{3^k}$ and $\tau^{v(j)}s - (\tau^{v(j)} - 1) 
\equiv s-r+1 \mod{3^k}$, and the first congruence implies  $(\tau^{v(j)} + 1)r \equiv  \tau^{v(j)}+1 \mod{3^k}$.
This gives $r \equiv 1 \mod 3^k$, which is impossible.

As in the proofs of Lemmas \ref{lemma52} and \ref{lemma53}, Cases C1 and C2 reduce to cases B2 and
B1, respectively, so both lead to contradictions. 
\end{proof}

\begin{lemma}
 \label{lemma55}
Suppose $g_k = (1, r, s)$ with $s > r$ is a function in $3^k$ variables and
$g_{k-1} = (1, [r/3] + 1, [s/3] + 1)$ is a function in $3^{k-1}$ variables.
Then $g_k$ belongs to an equivalence class of size $v(j) = 2 \cdot 3^j$ for some $j$, $1 \leq j \leq k-2$,
if and only if $g_{k-1}$ belongs to an equivalence class of the same size $v(j)$.
\end{lemma}

\begin{proof}
 First we suppose $g_k$ has $3^k$ variables and belongs to an equivalence class of size $v(j)$.  
By Lemma \ref{lemma54}, \eqref{eqn31} and \eqref{eqn32} hold for the given $j$, so it follows that 
if $\tau \not\equiv 0 \mod{3}$, then
\begin{equation}
\label{eqn37}
(\tau^{v(j)}-1)(r-1)/3 \equiv 0\mod{3^{k-1}} ~\text{and}~(\tau^{v(j)}-1)(s-1)/3 \equiv 0\mod{3^{k-1}}
\end{equation}
Clearly \eqref{eqn37} is equivalent to saying that if $\tau \not\equiv 0 \mod 3$, then  the function $h_{k-1} = (1, (r+2)/3, (s+2)/3)$ in
$3^{k-1}$ variables satisfies 
\begin{align}
\label{eqn38}
&\sigma_\tau^{v(j)}(1,(r+2)/3,(s+2)/3) =\\
\notag&(1,\tau^{v(j)}(r-1)/3 + 1, \tau^{v(j)}(s-1)/3 + 1) = (1, (r+2)/3,(s+2)/3) 
\end{align}
(the second equality follows from \eqref{eqn37}).  Since  $h_{k-1} = g_{k-1}$  by \eqref{eqn32}, \eqref{eqn38}
shows that the equivalence class of $g_{k-1}$ has size $\leq v(j)$.  If \eqref{eqn38} were true with $v(j)$ 
replaced by some proper divisor of $v(j)$, then \eqref{eqn37} would also be true with that replacement, 
contradicting our hypothesis that the class of $g_k$ has size $v(j)$.  Thus the class of $g_{k-1}$ must have size $v(j)$. 

For the converse, we assume $g_{k-1}$ in $3^{k-1}$ variables belongs to an equivalence class of size $v(j)$, $1 \leq j \leq k-2$.
Then reversing the above argument shows that $g_k$ in $3^k$ variables belongs to an equivalence class of the same size $v(j)$.
\end{proof}

\begin{lemma}
 \label{lemma56}
An equivalence class of size $2 \cdot 3^j$, $j \geq 1$, can occur only if 
$n \geq 3^{j+1}$.  If $U(j, k)$ denotes the number of equivalence classes of size  $2 \cdot 3^j$
when $n = 3^k$, then $U(j, k) = U(j, j+1)$ for all $k \geq j+1$.
\end{lemma}

\begin{proof}
The first sentence of the lemma follows from the fact that if $n = 3^k$, then the size of any equivalence
class must be a divisor of $|G_n| = \phi(3^k) = 2 \cdot 3^{k-1}$.  If there are $U(j, j+1)$ equivalence classes
of size $2 \cdot 3^j$ when $n = 3^k$, then the proofs of Lemmas \ref{lemma54} and \ref{lemma55} show that there 
are exactly $U(j, j+1)$ equivalence classes of size $2 \cdot 3^j$ when $n = 3^k$ for any $k \geq j + 1$. 
\end{proof}

Now we can prove the theorem which provides a detailed description of the equivalence classes. 

\begin{theorem}
\label{thm57}
Suppose $n = 3^k$ for $k \geq 1$. Then
\begin{equation} 
\label{eqn39}
E(3^k) = 3^{k-1}. 
\end{equation}
There is at least one equivalence class of size $d$ for every divisor $d \neq 2$ of $\phi(n) = 2 \cdot 3^{k-1}$,
and no class of size $2$.  If $d = 3^j$, $0 \leq j \leq k-1$, there is exactly one equivalence class of size $3^j$.
The least representative in the lexicographical ordering of $D_n$ for the class of size $3^j$ is 
$f_{j+1} = (1, 3^{k-j-1}+1, 2 \cdot 3^{k-j-1}+1)$ for $0 \leq j \leq k-1$.  If $d = 2 \cdot 3^j$, $1 \leq j \leq k-1$,
there are exactly $2 \cdot 3^{j-1}-1$ equivalence classes of size $2 \cdot 3^j$.
\end{theorem}

\begin{proof}
Let $F(n) = |D_n| =$ number of distinct cubic MRS functions $(1,j,k)$ in $n$ variables.
It follows from Lemma \ref{lem3} that $F(n) = (n^2 - 3n + 6)/6$ for all $n$ divisible by $3$. The assertions
that there is exactly one equivalence class of size $3^j$, $0 \leq j \leq k-1$, and that this class has least 
representative $f_{j+1}$ follow from Lemmas \ref{lemma51} and \ref{lemma53}.  The assertion that there is no 
class of size 2 is Lemma \ref{lemma52}. 
     
The case $n = 3$ is trivial (one equivalence class ${f_1} = {\;(1,2,3)\;}$) and the cases $n = 9$ and $27$ 
of the theorem are shown in Tables 1 and 2.  We assume that we have all the assertions in the theorem for all 
$n = 3^{j-1}$, $2 \leq j \leq k$, and we prove the theorem for the case $n = 3^k$.  We already have the cases 
for $k \leq 3$. The induction hypothesis proves the last sentence in the theorem for all $j \leq k-2$. 
     
By our induction hypothesis and Lemma \ref{lemma56}, for $n = 3^k$ there are $F(3^{k-1})$ functions in 
equivalence classes with size $\leq \phi(3^{k-1}) = 2 \cdot 3^{k-2}$.  By Lemma \ref{lemma53}, for $n = 3^k$
there is also exactly one equivalence class of size $3^{k-1}$.  The number of functions that are not in any of 
these classes for $n = 3^k$ is  $F(3^k) - F(3^{k-1}) - 3^{k-1} = (4 \cdot 3^{2k-2} - 6 \cdot 3^{k-1})/3 = L(k)$, say.
Since the size of every equivalence class divides $\phi(n) = 2 \cdot 3^{k-1}$, all the remaining equivalence classes must have size $2 \cdot 3^{k-1}$.  Thus there are exactly 
$L(k)/(2 \cdot 3^{k-1}) =  2 \cdot 3^{k-2} - 1$
classes of size $2 \cdot 3^{k-1}$. This proves the remaining case $j = k-1$ of the last sentence in the theorem,
and also completes the proof that there is at least one equivalence class of size $d$ for every divisor $d \neq 2$ of 
$\phi(n) = 2 \cdot 3^{k-1}$.  Also we see that $U(j, j+1)$ in Lemma \ref{lemma56} is $2 \cdot 3^{j-1} - 1$ for  $j \geq 1$.

The previous work shows that the number $E(3^k)$ of equivalence classes is 
\begin{equation*}
            \sum_{j=0}^{k-1}  1  +  \sum_{j=1}^{k-1}  (2 \cdot 3^{j-1} - 1) = 3^{k-1}, 
\end{equation*}
which proves \eqref{eqn39}.
\end{proof}
\begin{remark} 
It should be possible to extend the techniques in this paper to give a detailed description, similar to the one in Theorem \ref{thm57}, for the cases when $n$ is a power of any prime.  The case $n = 2^k$ will require special treatment.  We do not consider these extensions in this paper.
\end{remark}
\begin{remark} 
I acknowledge the valuable contributions of my research assistants Max Bileschi and Dan Padgett, whose work greatly facilitated the writing of this paper.  Special thanks are due to Dan Padgett for his proof of Lemma \ref{lem6}, which is simpler than my original proof.  I am grateful to Yuri Borissov for telling me about the Online Database of Boolean Functions \cite{ODBF}. 
\end{remark}

\pagebreak

\end{document}